\documentclass[a4paper, 12pt]{article}
\usepackage{setspace}
\usepackage[margin=2.5cm]{geometry}
\usepackage{authblk}
\usepackage[utf8]{inputenc}
\usepackage{threeparttable}
\usepackage{booktabs}
\usepackage{adjustbox}
\usepackage{tabularx}
\usepackage{amssymb}
\usepackage{amsfonts}
\usepackage{amsmath}
\usepackage{natbib}
\usepackage{csquotes}
\usepackage[bookmarks=false, pdfstartview={FitH}, colorlinks=true, hidelinks, urlbordercolor={1 1 1}]{hyperref} 
\usepackage{xcolor}
\usepackage{appendix}
\usepackage{multirow}

\author{Mario A. Maggioni and Domenico Rossignoli}
\affil{\textit{\normalsize CSCC, HuRoLab, and DISEIS, Universit\`a Cattolica del Sacro Cuore, Milano}}

\title{If it Looks like a Human and Speaks like a Human ... \\
{\large Dialogue and cooperation in human-robot interactions}}

\date{This version: \today}

\begin{document}

\maketitle

\begin{abstract}
\noindent
This paper presents the results of a behavioral experiment conducted between February 2020 and March 2021 at Universit\`a Cattolica del Sacro Cuore, Milan Campus, in which students were matched with either a human or a humanoid robotic partner to play an iterated Prisoner’s Dilemma. The results of a Logit estimation procedure show that subjects are more likely to cooperate with human rather than with robotic partners; that they are more likely to cooperate after receiving a dialogic verbal reaction following a sub-optimal social outcome; and that the effect of the verbal reaction is not dependent on the nature of the partner. Our findings provide new evidence on the effects of verbal communication in strategic frameworks. The results are robust to the exclusion of students of Economics-related subjects, to the inclusion of a set of psychological and behavioral controls, to the way subjects perceive robots’ behavior, and to potential gender biases in human-human interactions.
\end{abstract}

\medskip


\vskip 50pt

\noindent\textbf{Acknowledgments}

\noindent We acknowledge funding from two competitive research funds of Universit\`a Cattolica del Sacro Cuore: D3.2/2018 ``Human-Robot confluence'' and D3.2/2020 ``Behavioral change''. This project is a spin-off of the Fetzer Institute's ``\#3534.00'' project granted to M.A. Maggioni  on the use of behavioral economics techniques to assess behavioral change.
We thank M. Colagrossi, P. Natale, C. March, L. Stella, F. Trombetta, and F. Perali for useful comments and observations on previous versions of this paper. The usual \textit{caveats} apply.  
Research assistantship by E. Cerolini, P. Gambacciani, C. Marconi, F. Manzi and P. Zaza is acknowledged. Special mention is due to F. Manzi whose expertise and skills were precious for programming the robot in the lab experiment. We also thank N. Graverini and P. Alberti for technical support.

\setstretch{1.5}

\clearpage

\section{Introduction}

For decades human-robot interactions have played an important part in the production of cultural products (stories, novels, movies or, more recently, videogames), populating the imagination of people across countries and cultures. More recently, we have witnessed anthropomorphic robots becoming more widespread beyond movie sets or research labs, paving the way for them to play an increasing role in economic and social interactions. 

These robots move and, most importantly, talk somehow like human beings; this last feature is particularly relevant, since the ability to communicate through words constitutes a distinctive feature of human interactions.
A long tradition in Philosophy, dating back to Socrates (469–399 BC), has stressed the ontological foundations of dialogic relationships. Martin \cite{buber58and} claimed that full understanding of one's own identity is strictly dependent on dialogue with another presence. 

More recently, \citet{dumouchel2017living} and \citet{damiano2018anthropomorphism} have argued that dialogue is the fundamental structure and basic pattern of humans' actions and thinking. 
Paleoanthropologists such as 
\cite{dunbar1996grooming}, \cite{lieberman2006toward} and \cite{tattersall2008evolutionary} as well as paleoeconomists such as \cite{HORAN20051,horan2008coevolution} have formulated the hypothesis that verbal interactions and \enquote{verbal grooming} were the crucial factors in the emergence of the evolutionary advantage  of \textit{Homo Sapiens} over \textit{Homo Neanderthalensis}, (thus explaining the survival of the former) some 150 thousand years ago; thus dialogue has been somehow embedded in the evolutionary process of human nature.\footnote{For a recent, currently isolated, dissenting voice, see \citet{conde2021neanderthals}.}

Further, communication has been empirically proven to increase trust and cooperation \citep{balliet2010communication}, which in turn are pivotal in fostering positive social interactions. In particular, face-to-face communication has been shown to both promote and sustain cooperation between subjects even in strategic settings such as social dilemmas \citep{ostrom1997neither, ostrom2000collective,bicchieri2002covenants}. 

In recent decades, the dialogic framework has been enlarged in scope by the increase in the use of anthropomorphic robots and by the development of the sub-fields of Social Robotics (SR) and Human-Robot Interactions (HRI), which study mechanical objects that are able to communicate, both verbally and non-verbally, in similar ways to human beings and to play the role of artificial subjects, acting as ``social partners'' \citep{fong2003survey,sung2010domestic,dumouchel2017living}.
Recent empirical research in HRI has implemented experimental frameworks in which human subjects have been partnered with humanoid robots in social dilemmas \citep[see, among others,][]{krach2008can,de2011effect,desteno2012detecting,paeng2016human,de2020interplay}. 
This process has almost seamlessly led on to the question of whether communication also promotes cooperation in these new settings where people interact with robotic agents.


To address this important issue in social science, in this paper, we devised a randomized experiment in which human subjects are randomly matched with either a human or an anthropomorphic robot partner (NAO, produced by Softbank Robotics) and asked to perform an iterated Prisoner's Dilemma (PD). In each of these two sub-samples, after the first round of the game, and prior to a second round being proposed, half of the subjects are randomly assigned to treatment, which consisted of the partner providing a \textit{Dialogic Verbal Reaction} (henceforth, DVR) if a sub-optimal social outcome had occurred in the first round of the PD. 

The aim of our experiment is threefold: first, to investigate whether subjects behave differently depending on the nature of their partner (whether human or robot); second, to analyze whether a DVR, that implicitly refers to cooperation as a socially desirable strategy, influences the subsequent choice of the subject; third, and most importantly, to check whether the effect of this verbal reaction depends on the (human or robotic) nature of the partner.\footnote{In a sense, our third research question can be thought of as a sort of modified Turing Test. In the original three-person ``imitation game'' \citep{machinery1950computing} an interrogator chats with two respondents, located in separate rooms, asking questions to detect which one of the two is a machine. If the interrogator cannot reliably tell the machine from the human, the machine is said to have passed the test. In our experiment the subject knows that the robot is a machine but he/she is surprised by the DVR and reacts as if the robot were a human.}

Our main result shows that being exposed to a DVR, following sub-optimal social outcomes, positively affects the cooperation rate of subjects at the next stage and that this effect is not dependent on the partner's type. 
In other words, despite the fact that no DVRs affect the players' payoffs, as long as the partner subtly evokes cooperation - either in the form of an apology or a reprimand or disappointment - the subject's subsequent choice tends towards cooperation, irrespective of whether their partner is a fellow human being or an anthropomorphic robot. 

This paper contributes to the existing economic literature in two ways. Firstly, we provide new evidence on the effectiveness of communication in affecting decision-making, extending this result to HRI; secondly, and most importantly, we show that the effect of a DVR is independent of the human vs artificial nature of the agent, thus contributing to an area in the study of economic interactions between human and non-human subjects that remains under-researched.

\section{Related literature}

Several empirical works \citep[e.g.][]{dawes1977behavior, braver1986choices, ostrom1991communication, bicchieri2002covenants} have tested the importance of communication in influencing the outcome of strategic interactions. \cite{sally1995conversation} published the first meta-analysis of this stream of literature and concluded that communication exerts the strongest effect, relative to other variables known to influence cooperation, such as group size, the magnitude of the reward for choosing not to cooperate, and group identity. Subsequently, \cite{balliet2010communication} addressed the same issue through an improved meta-analysis, by adopting mediation-analysis techniques, and confirmed that communication has a strong positive effect on cooperation within social dilemmas. 

In a different epistemological tradition, ``critical social theory'', Habermas 
relies on the notion of ``strategic competence'' in his account of the development of moral consciousness, defined as ``the ability to make use of interactive competence for consciously processing morally relevant conflicts of action'' \citep[p.88]{habermas1979communication}. He also refers to the ability of social actors to examine the validity and legitimacy of established norms discursively. 
Habermas thus \enquote{identifies a mechanism that might compellingly account for the binding force of language in strategic interaction} \cite[p. 81]{johnson1993talk}.


The nature of inter-subject interactions has lately been extended to different types of agents by the more widespread use of robots and, specifically, of humanoid social robots. The original goal of robotics was to create ``artificial workers'' engaged in a broad range of activities.\footnote{Such as, among others, production, information, education, coaching, therapeutic mediation, assistance, entertainment, and companionship.} However, computer scientists and engineers soon had to acknowledge that, to be able to operate in many of these fields, robots needed to exhibit a variety of social behaviors and, in particular, to evince a believable ``social presence'', defined as a robot’s capability to give the user the \enquote{sense of being with another} \citep{biocca2003toward}, or the \enquote{feeling of being in the company of someone} \citep{heerink2008influence}. For this reason, social robots have been developed to express and/or perceive emotions; communicate using high-level dialogue; and learn/recognize models of other agents; establish/maintain social relationships; use natural cues (gaze, gestures, etc.); exhibit distinctive personality and character; learn/develop social competencies.\footnote{Indeed, a number of recent studies have investigated the potential effects of robots on human behavior. For example, on the one hand there is increasing concern that the complex relationship between humans and machines may have detrimental effects on the mental health of workers \citep{robelski2018human} and might function as an additional stressor in the workplace \citep{korner2019perceived}. On the other hand, in some cases interactions with robots have been shown to be perceived as skill enhancing and capable of increasing job satisfaction. \citep{compagni2015early} and have shown that the social perception of robots is influenced by physical appearance and behavior \citep{marchetti2018theory, manzi2021can}.}

The growth in the use of humanoid robots and the emergence of the scientific sub-fields of SR and HRI, has spurred interest in analyzing the possible consequences of repeated interactions between experimental subjects and ``social robots'' i.e. artificial agents that exhibit one or more of the previously defined ``social'' characteristics \citep{fong2003survey,dumouchel2017living,Gaggioli2021relat}. 

In economic literature, the decision to trust another agent to behave as a trustworthy partner in a transaction or to cooperate in a social dilemma\footnote{Commonly operationalized within a game-theoretical framework through a Prisoner's Dilemma, a Trust Game, a Centipede Game, or a ``lost letter'' experiment. See, among others, \cite{,dasgupta1988trust}, \cite{kreps1990micro} \cite{yezer1996does} and \cite{skeath1999games}.} is generally seen as not consistent with the pursuit of individual self-interest. For this reason, these decisions of trust are usually explained either in terms of repeated interactions within a finite uncertain time horizon, or by assuming: agents endowed with ``self-regarding preferences'', an  appropriate time discount rate, a certain degree of uncertainty over the type of the opponent, a ``warm glow" effect'' \citep{andreoni1990impure}, a ``gift-giving'' behavior \citep{akerlof1982labor}; or, finally, by referring to concepts such as equity and fairness \citep{fehr1999theory, rabin1993incorporating} in case of agents endowed with ``other-regarding preferences''. 

Relatively few papers explicitly consider the real relational dimension of agents' interactions in social dilemmas. In those papers, cooperative and trustful behaviors are explained as being motivated by an acknowledgment of the other party's attitudes and intentions. Similarly, the literature on so-called psychological games addresses the role of subjects' intentions, by making payoffs belief-dependent \citep{geanakoplos1989psychological,rabin1993incorporating, dufwenberg2004theory,deangelo2020psychological}. 

Finally, support for the importance of attitudes, intentions and verbal and non-verbal cues in communication emerges from behavioral economics, as well as psychology and neuroeconomics experiments, where players show different behaviors and neurological activations when playing incentivized tasks and games with human counterparts as opposed to \textit{automata} (ranging from PCs to robots with various degrees of humanization), despite facing identical material payoffs \citep{kiesler1996prisoner,mccabe2001functional,rilling2002neural,rilling2004neural,bicchieri2007computer,krach2008can,miwa2012impact,de2011effect,nouri2013cross,paeng2016human,wu2016trust,terada2017emotional,crandall2018cooperating,ishowo2019behavioural}. 

Among studies involving humanoid robots, the use of NAO has become increasingly popular due to the robot's features and capabilities, that make it appropriate for experimental (especially clinical) researches. \citet{robaczewski2020socially} documents 70 experimental studies in which subjects are involved in a HRI with NAO. In particular, 26 studies are specifically designed to study social interactions between human and robots, and only 5 of them especially focus on communication as the main moderating feature. To the best of our knowledge, only one of these studies \citep[i.e.][]{sandoval2016reciprocity} ask participants to play a Prisoner's Dilemma - and an Ultimatum Game - and none of them is designed to infer the effects of verbal messages evoking social norms to subjects' behavior.\footnote{On the contrary, in \citet{sandoval2016reciprocity} human partners were asked to be neutral, to interact as little as possible with participants and to avoid conversations. They were even instructed to nod at participants in return for their greetings at the beginning of the experiment. Further, an experimenter (called ``referee'') was always present during the interactions in the lab room, thus likely introducing a strong bias in participants' behavior.}




\section{Research design}
Our research design addresses three specific questions that relates to the way humans interact with humanoid robots in a repeated Prisoner's Dilemma.

\paragraph{RQ1:} \textbf{Do subjects' cooperation rates differ according to partner types? }
Following recent developments in empirical research in economics, psychology and social robotics we aim to understand whether subjects with a robotic as opposed to a human partner choose to cooperate with different or indistinguishable probabilities.

\paragraph{RQ2:} \textbf{Does a DVR affects the subjects subsequent behavior (irrespective of partner type) in terms of cooperation rates?}
Empirical evidence in the literature on social dilemmas shows that communication tends to promote cooperation. We aim to investigate whether subjects are more likely to cooperate after the partner has activated a DVR, after observing a sub-optimal outcome in the previous round of the game. 

\paragraph{RQ3:} \textbf{Are subjects' reactions to a DVR dependent on the partner's nature (human vs robot)?} This question originates directly from the former two. Since our experiment is designed as a 2$\times$2 matrix (see Table \ref{tab:expdesign} below) we may expect subjects' decisions to be affected affected by either the nature of the partner or a DVR (or both). In other words, we want to explore whether any of these effects is characterized by heterogeneity.

\medskip

To investigate how subjects behave when faced with a robotic partner we needed to exclude their initial (online) choice in Phase 1 to cooperate or not to cooperate to be dependent on the partner's type. At the same time, we could not entirely rely on anonymized interactions, since the purpose of the investigation is strictly related to the possible effects arising from the interactions with different types of partner.

For this reason, we designed an experimental procedure with two distinct phases. In Phase 1, we asked the subjects (university students) to answer an online questionnaire and to play an incentivized task against an unknown anonymous partner. At the end of Phase 1 we asked subjects whether they wanted to come to the University Lab, proceeding to Phase 2, in order to: (i) learn the result of the interaction; (ii) be rewarded; and (iii) possibly have other interactions with the partner.

\begin{table}[h]
\centering
\caption{Experimental design}
\begin{threeparttable}
\begin{tabular}{clll}
\toprule
                                            &                   &    \multicolumn{2}{c}{Treatment group} \\ \cmidrule(lr){3-4}
                                            &                   &    \textbf{No DVR}    & \textbf{DVR}         \\ \midrule
\multirow{2.5}{6em}{Experimental condition} &    \textbf{Human} & Baseline                              & Reaction    \\ [1ex]
                                            &    \textbf{Robot} & Robot                                 & Interaction \\
\bottomrule
\end{tabular}
\label{tab:expdesign}
\end{threeparttable}
\end{table}

We devised a $2\times2$ experimental design, as summarized in Table \ref{tab:expdesign}: on the one hand, we randomly assigned subjects to be matched with either a Human (H) or Robot (R) partner in the interactive situation (the Prisoner's Dilemma); on the other hand, we randomly administered a stimulus (treatment) to a fraction of our subjects in both the H and R experimental conditions. 

The treatment consisted of a ``Dialogic Verbal Reaction'' that the partner delivers after observing a sub-optimal aggregate outcome of the interaction. Different stimuli were administered depending on the observed outcomes in the Prisoner's Dilemma,\footnote{No DVR is activated when the aggregate Pareto optimal outcome is obtained, i.e when both subject and partner cooperate.} as summarized in Table \ref{tab:stimulus} in \ref{app:addtab}.

\subsection{Experimental procedures}
\begin{figure}
    \centering
    \frame{\includegraphics[width=1\textwidth,keepaspectratio]{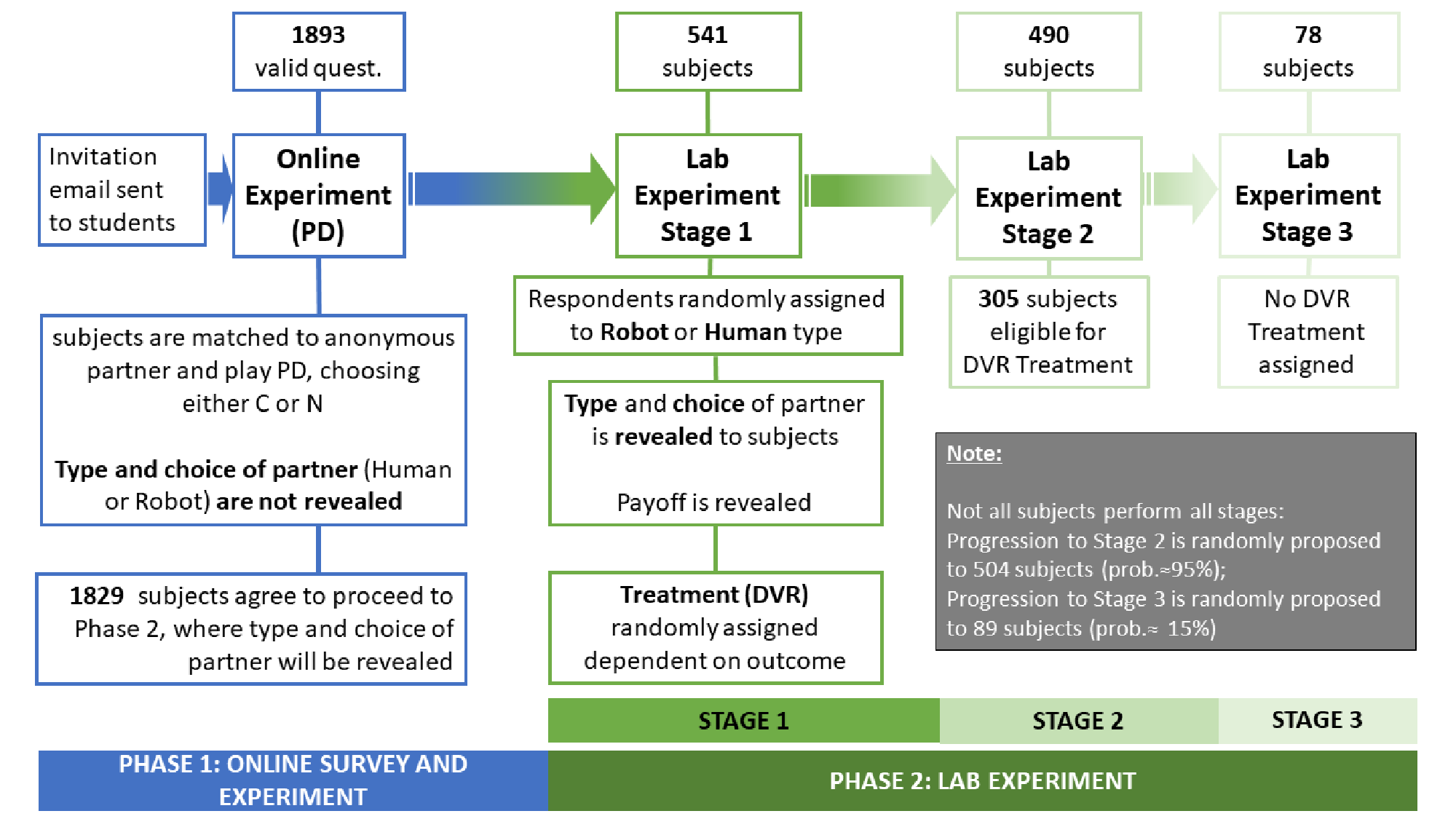}}
    \caption{Flow chart of experiment. Blue and green colors for boxes and frames refer respectively to Phase 1 (Online) and Phase 2 (Lab) of the experiment.}
    \label{fig:experimentprocedure}
\end{figure}

The layout of our experiment is summarized by the flow-chart in Figure \ref{fig:experimentprocedure}.


We sent invitation emails to all freshmen and sophomore students of \textit{Universit\`a Cattolica del Sacro Cuore}, who were attending lectures at the Milan campus.\footnote{Università Cattolica del Sacro Cuore is a multi-campus university mainly based in Milan but with other locations in Rome, Brescia, Piacenza and Cremona.}
The email included a link to an online survey containing the first phase of the experiment. In addition to a set of psychological, attitudinal and socio-demographic questions, the survey included the first round of a repeated Prisoner's Dilemma \citep[as in][]{kreps1982rational}. At the beginning of each round, each player is given 3 euros - as initial endowment - and can choose between two actions: if he/she chooses to ``Cooperate'', their sum is transferred to their Partner who receives double the original sum. If he/she chooses ``Not Cooperate'', he/she keeps their original sum. Thus, the game proposed to subjects in this experiment consists of N repetitions of a two-person, two-strategy, game (as summarized in Table \ref{tab:payoff}). 

\begin{table}[h]
\centering
\caption{Experiment's monetary payoff matrix}
\begin{threeparttable}
\begin{tabular}{llcc}
\toprule
  &                               &               \multicolumn{2}{c}{Partner's choice} \\ \cmidrule(lr){3-4}
 &                               & \textbf{Cooperate (C)}    & \textbf{Not Cooperate (N)}         \\ \midrule
\multirow{2.5}{4em}{Subject's choice} & \textbf{Cooperate (C)}        & (6;6)                     & (0;9)    \\ [1ex]
                                    & \textbf{Not Cooperate (N)}    & (9;0)                     & (3;3) \\ 
                                    \bottomrule
\end{tabular}
\label{tab:payoff}
\end{threeparttable}
\end{table}

Players, in each round, choose simultaneously. At the end of each round, players are reminded of their own and their opponents' choices, and shown the resulting payoffs. Total payoffs are the undiscounted sums of the round payoffs (plus the one-off show-up fee).\footnote{It seems reasonable to assume that IDR is not relevant for our experimental setting given that: (i) when subjects play their first round they are unaware of the delay between the first and second rounds; (ii) this delay is short (on average about 7 days); (iii) all subsequent rounds are played immediately one after the other in the experiment room. However, to address  potential biases driven by this issue, in Section \ref{sec:robustness} we also control for a measure of intertemporal discount rate.}

Students who agreed to take part in the incentivized game were informed that they had been randomly paired with an unknown anonymous (human or artificial) partner who was playing the same game.\footnote{The partners' cooperation rate was set at 50\%, to be consistent with the average cooperation rate observed in 8 experimental studies, collected in 5 papers, involving Italian subjects, namely \citet{Pepitone_1967, Pepitone_1970, gallucci2000experimental, Ciardo_2015, Meier_2016}. The weighted average of the cooperation rate in these studies is 49.23\%. Data retrieved from \url{https://app.cooperationdatabank.org/}.} If they agreed to participate, they were asked to make their choice (either Cooperate, C, or Not cooperate, N). The outcome of the experimental session (and the monetary reward, including show-up fee) would only be revealed on participants' agreeing to attend and take part (in person) in the second phase of the experiment at the university Lab. Only at that point would respondents discover whether the partner they had faced in the online interaction was a human or a robot, the outcome of the interaction would be revealed and they might receive an invitation to play one (or more) rounds of the game again.

Students who agreed to take part in Phase 2 had to schedule an appointment through an online third-party application. Phase 2, the proper Lab Experiment, consisted of three sequential stages. In Stage 1, the subject was taken to the lab room by the experimenter, where he/she could meet and face his/her partner, whether Human or Robot,\footnote{The Robot was NAO, a humanoid robot produced by Softbank Robotics (see Figure \ref{fig:nao} in \ref{app:addfig} for a picture, \cite{Gelin2018} and \cite{robaczewski2020socially} for references); human partners were Ph.D. students recruited and trained for this task.} with whom he/she was told he/she had played the PD online in Phase 1. When the subject entered the experiment room, the robot was already there, whereas the human partner entered the room from the control room.


\begin{figure}
    \centering
    \frame{\includegraphics[width=1\textwidth,keepaspectratio]{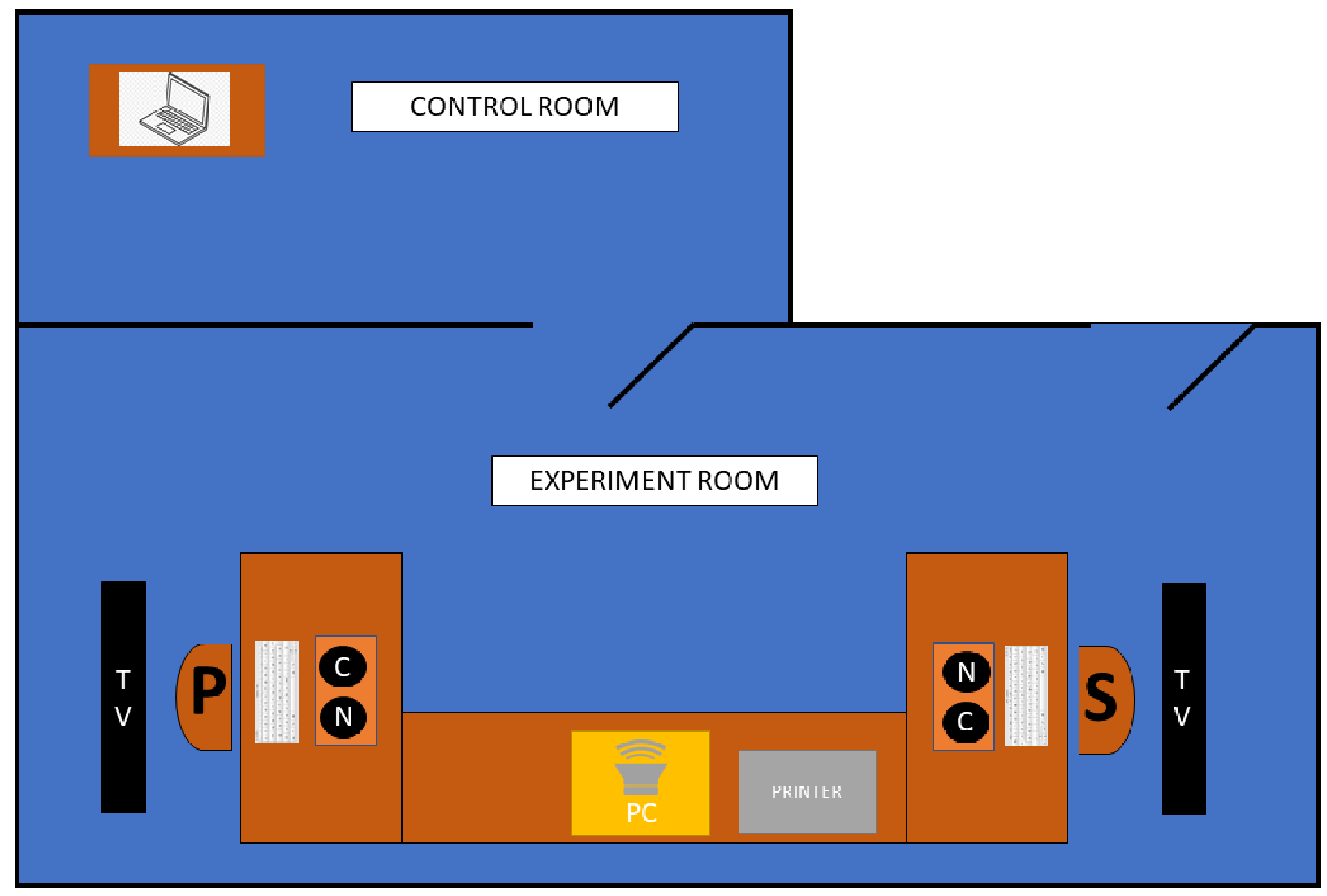}}
    \caption{Experiment room set}
    \label{fig:roomset}
\end{figure}

Having given the subject instructions about the experimental procedures, the experimenter would tell the players to wait and listen to the game director's instructions which were available both as a recorded voice message as well as being displayed on the screen in front of each player; then the experimenter left the room. The subject would sit at a table, then the partner (either a Ph.D. student or a humanoid robot) would greet the subject and verbally introduce him/her/itself.\footnote{In all sessions, the Ph.D. students introduced themselves by their first name, greeted the subjects and asked the subject's first name; the robot, introduced itself stating the following in all experimental sessions: ``Hi, I am ToM, a humanoid robot developed by Softbank Robotics able to perform complex interactions with human beings. What's your name?''} Finally, the screen would reveal the outcome of the Prisoner's Dilemma played online in Phase 1. 

Based on the game outcome as shown in Table \ref{tab:stimulus} in \ref{app:addtab}, a random algorithm determined whether to activate the DVR with a 50\% probability, triggering the partner, whether human or robot, to deliver the appropriate verbal stimulus.

The subject was then asked whether he/she would like to continue to play another round of the game, according to a random algorithm with a probability of 95\%. If the subject accepted, a second round was then implemented, without any further assignment of treatment.We checked that the probability of both proposal and agreement to proceed to Stage 2 was not conditional on the outcome observed in Stage 1.\footnote{We analyzed the association between a categorical variable, identifying all the 4 possible outcomes observed in Stage 1, and (i) a binary variable, identifying whether Stage 2 had been proposed to the subject; (ii) a binary variable, identifying whether this proposal had been accepted. In both cases, we are able to exclude any non-random pattern. In case (i), a Chi-sq. test ($Chi=4.723, df=3, p=0.193$) and an ANOVA ($F-stat=1.58, p=0.194$) allows to infer that the proposal to continue to Stage 2 was independent of the observed outcome of Stage 1. Figure \ref{fig:bar_gostage2} in \ref{app:addfig} summarizes this outcome; in case (ii) the result is even stronger since only 16 subjects out of 506 (96.8\%) refused to continue, and any association between acceptance and outcome in Stage 1 could be excluded by both a Chi-sq. test ($Chi=1.111, df=3, p=0.774$) and an ANOVA ($F-stat=0.37, p=0.776$). Figure \ref{fig:bar_acceptstage2} summarizes this result.}

After the results of the second round were revealed, the subject could ask to continue with a further round of the game, according to a random algorithm with a 15\% probability. If the subject agreed to play again, a third round of a PD was implemented, again without any treatment assigned.\footnote{Stage 3 of the experiment was of no interest for the analysis. However, we devised the possibility of playing 3 rounds of the PD to generate uncertainty about the total number of rounds in case of information sharing between students.}

After the last incentivized interaction, the game director communicated the total amount gained and asked the subject whether he/she would like to donate 1 euro to a charity of his/her choice.\footnote{The available alternatives were \textit{M\'edecins Sans Fronti\`eres} and \textit{Greenpeace}.}

Once this last choice was made, the subject had his/her receipt automatically printed, he/she left the lab room to enter the check-out area where he/she was paid the amount in cash. On leaving the Lab the student signed a confidentiality agreement requiring him/her not to share any information about the experiment with fellow students.

We planned to run the experiment from 13\textsuperscript{th} February, to 13\textsuperscript{th} March, 2020. However, due to the outbreak of the COVID-19 pandemic, we had to stop all activities in the lab on 21\textsuperscript{th} February, which proved to be our last day of data collection in 2020. 

In February 2021, universities in Italy opened again for in-person lectures. We seized the moment\footnote{Invitations were sent to freshmen and, to avoid duplication to those sophomores who had not opened the invitation e-mail we sent them when they were freshmen in 2020.} and intended to run another wave of the entire experiment (both Phases 1 and 2) between 22\textsuperscript{nd} February and 31\textsuperscript{th} March. Again, due to the resurgence of the pandemic, we had to stop all activities in the lab on 4\textsuperscript{th} March. 
At the end of the experimental session we fully disclosed all experimental procedures to all participants.

\section{Data and estimation methods}
\subsection{The sample}
We sent a total of 23,552 emails, and received 2,205 individual answers and 1,893 valid and completed questionnaires (Phase 1). 1,829 respondents (96.6\% of the valid entries) agreed to participate in the Second Phase of the experiment.

Out of a total of 490 subjects taking part in Phase 2 (see table \ref{tab:sumstat}, bottom panel), the sample eligible for an analysis of the treatment effect consists of 305 subjects who had made their choice in Round 2 having been assigned either to the Treatment group (Dialogic Verbal Reaction, DVR) or to the Control group (No DVR) (see table \ref{tab:sumstat}, top panel).\footnote{As illustrated in Table \ref{tab:stimulus} in \ref{app:addtab} a DVR stimulus could only be applied in 3 out of 4 possible outcomes.} An initial inspection shows that subjects taking part in the research were overall more likely to cooperate than not to cooperate.\footnote{Please note that most of the cooperative choices, i.e. those yielding an outcome of ``CC'' in Stage 1, were not included in Stage 2 in the Experimental sample: therefore, by design, the Experimental sample is ``less cooperative'' than the Full sample.} 

Table \ref{tab:sumstatTREAT} shows the summary statistics of both the outcome and control variables by treatment group and experimental condition: the Robot and Human subsamples are shown in the top and bottom panels respectively; treatment and control values are shown across the columns. 
As the table shows, all subsamples are well balanced across treatment groups in terms of control variables. The only imbalance relates to Female, which is less represented in the No DVR group in the Robot condition (top panel) and more represented in the same group in the Human condition. A summary of balance tests, estimated using a Logit model, is shown in Figure \ref{fig:balanceTREAT} displayed in \ref{app:balance}, while Figure \ref{fig:balancePARTNER} reports the coefficients of balance tests including for partner types.\footnote{In this case, Freshman is statistically significant, and greater in the case of Robot rather than Human partner. In Figure \ref{fig:balanceWAVE} we also provide a balance test for the two experimental waves, 2020 and 2021, showing that only Freshmen are more represented in 2021.}

Interestingly, also the baseline measure of our outcome variable (i.e. the choice to Cooperate in Stage 1) is not statistically different across treatment groups.  Conversely, a t-test shows that in both samples subjects assigned to the treatment group are more likely to cooperate than those in the control group, suggesting a potential effect of the DVR on subjects' choices in Stage 2.

Finally, the bottom panel of Table \ref{tab:sumstatTREAT} provides a summary of the differences in DVR effects by subsamples, i.e. it shows whether the increase in cooperation in the treatment group is heterogeneous across Human/Robot condition. Interestingly, being assigned to the Human condition is likely to increase the probability of cooperation in Stage 2 in the control group, where no DVR is performed; however, this statistical difference disappears when the comparison is made in the treatment group, suggesting that the effect of DVR is not heterogeneous across partner types.

\begin{table}[h]\centering
\def\sym#1{\ifmmode^{#1}\else\(^{#1}\)\fi}
\caption{Summary statistics, by experimental conditions}
\begin{adjustbox}{width=1\textwidth}
\centering
\begin{threeparttable}
\begin{tabular}{l*{3}{ccccc}}
\toprule
\textbf{Partner=Robot} &&&&&&&& \\
                    &\multicolumn{3}{c}{DVR group}               &\multicolumn{3}{c}{No DVR (control group)} & \multicolumn{2}{c}{T-test} \\\cmidrule(lr){2-4}\cmidrule(lr){5-7}\cmidrule(lr){8-9}
            &        Mean& St. Dev.&  Obs &  Mean &   St. Dev & Obs &   Diff. & t-stat \\ \cmidrule(lr){2-9}
\rule{0pt}{4ex}\textit{Outcome variables}&            &            &            &            &            &            &                     &            \\
Cooperate (Stage 1) &       0.736&       0.443&          91&       0.650&       0.480&          80&      -0.086         &     (-1.22)\\
Cooperate (Stage 2) &       0.630&       0.486&          81&       0.380&       0.489&          71&      -0.249\sym{**} &     (-3.15)\\
\rule{0pt}{4ex}\textit{Control variables}&            &            &            &            &            &            &                     &            \\
Female              &       0.769&       0.424&          91&       0.575&       0.497&          80&      -0.194\sym{**} &     (-2.73)\\
Freshman            &       0.857&       0.352&          91&       0.800&       0.403&          80&      -0.057         &     (-0.98)\\
Economics           &       0.440&       0.499&          91&       0.450&       0.501&          80&       0.010         &      (0.14)\\
Test failed         &       0.099&       0.300&          91&       0.075&       0.265&          80&      -0.024         &     (-0.55)\\
\midrule
\textbf{Partner=Human} &&&&&&&& \\
                    &\multicolumn{3}{c}{DVR group}               &\multicolumn{3}{c}{No DVR (control group)} & \multicolumn{2}{c}{T-test} \\\cmidrule(lr){2-4}\cmidrule(lr){5-7}\cmidrule(lr){8-9}
            &        Mean& St. Dev.&  Obs &  Mean &   St. Dev & Obs &   Diff. & t-stat \\ \cmidrule(lr){2-9}
\rule{0pt}{4ex}\textit{Outcome variables}&            &            &            &            &            &            &                     &            \\
Cooperate (Stage 1) &       0.640&       0.483&          86&       0.663&       0.476&          86&       0.023         &      (0.32)\\
Cooperate (Stage 2) &       0.737&       0.443&          76&       0.558&       0.500&          77&      -0.178\sym{*}  &     (-2.34)\\
\rule{0pt}{4ex}\textit{Control variables}&            &            &            &            &            &            &                     &            \\
Female              &       0.628&       0.486&          86&       0.802&       0.401&          86&       0.174\sym{*}  &      (2.57)\\
Freshman            &       0.663&       0.476&          86&       0.791&       0.409&          86&       0.128         &      (1.89)\\
Economics           &       0.302&       0.462&          86&       0.395&       0.492&          86&       0.093         &      (1.28)\\
Test failed         &       0.151&       0.360&          86&       0.186&       0.391&          86&       0.035         &      (0.61)\\
\midrule
\textbf{Robot vs. Human comparison} &&&&&&&& \\
&&&&&& Obs & Diff. & t-stat \\ \cmidrule(lr){7-9}
Cooperate (Stage 2): DVR &&&&&& 157 & -0.11  & (1.44) \\
Cooperate (Stage 2): No DVR (control group) &&&&&& 148 & -0.18**  & (2.19) \\
\bottomrule
\end{tabular}
\begin{tablenotes}
\footnotesize
\item \textit{Notes:} Summary statistics refer to subjects eligible for treatment, i.e. excluding those yielding ``CC'' as the outcome of Stage 1. 
\end{tablenotes}
\end{threeparttable}
\end{adjustbox}
\label{tab:sumstatTREAT}
\end{table}

\subsection{Estimation technique}
We addressed the aforementioned set of research questions by implementing a Logit model, in which the probability that respondent $i$ makes a Cooperative choice is conditional on a set of control variables and experimental conditions. Formally:

\[ Y_{i,s} =
\begin{cases}
 1  & \quad \text{if respondent $i$ chooses Cooperation at round $r$} \\
 0  & \quad \text{if respondent $i$ chooses Non-Cooperation at round $r$}\\
\end{cases}
\]

\begin{equation}\begin{split}
\label{eq:2}
log\left(\frac{\pi_{i,r}}{1-\pi_{i,r}}\right) &= \beta_0 + \delta_1 Robot_i + \delta_2 DVR_i + \delta_3 Robot_i\times DVR_i \\
&+ \beta_1Choice_{i,{r-1}} + \beta_jX_{j,i} \cdots + \beta_kX_{k,i}
\end{split}
\end{equation}

where
$\pi_i$ is the probability that $Y_i$ equals 1; $Y$ is the dependent variable measured at rounds 1 and 2; $\beta_0, \beta_1, \beta_j, \cdots, \beta_k$ and $\delta$ are the parameters to be estimated, with $\delta_1, \delta_2$ and $\delta_3$ being respectively the main coefficients of interest for RQ1, RQ2 and RQ3;
$Choice_i,_{r-1}$ is the choice made by respondent $i$ in the previous experimental round, $r-1$; and $X_j \cdots X_k$ are a set of $k$ experiment-related and control variables, illustrated below.

To increase the precision of the estimates and to account for potential confounding factors, all models are estimated using different specifications. Control variables include experiment-related controls and background characteristics of the subjects.
In the former set of controls we include \textit{Instruction order}, a categorical variable that takes into account which of the potential outcomes of the Prisoner's Dilemma is shown first to the subject during the introductory instructions for the game, to control for potential priming;\footnote{Subjects received the illustration of all potential outcomes of PD in random order. We control for the outcome that appears first to the subject, to account for the instructions having a potential priming effect. We also replicated our results re-coding this variable to account for the outcome that was shown last. The results remain unchanged.} \textit{Wave}, to control for the year of the experiment (either 2020 or 2021); and \textit{Experiment day} which relates to the number of days the experiment has been running (e.g. Day 1, 2, 3, ...) to control for potential spillover effects.  
In the latter group of controls we include \textit{Female}, to control for subjects' gender; \textit{Freshman}, to control for subjects undertaking the first year of their BA; \textit{Econ}, to control for subjects enrolled in BA degrees in Economics, Management, Finance and Banking; and \textit{Fail test}, to control for subjects failing the pre-game test designed to verify their full understanding of the game's procedures, rules and payoffs. Moreover, all models are also estimated by excluding this latter group of subjects. The description of all variables is summarized in Table \ref{tab:descvars} in \ref{app:addtab}.

Finally, since our experiment design consists of $2\times2$ treatment conditions, in Section \ref{sec:robustness} we test the first two hypotheses by estimating separately the effects of $Robot$ and $DVR$ using sample splits. 

\section{Results}
\label{sec:results}
\subsection{Cooperation patterns}
\begin{figure}[h]
    \centering
    \includegraphics[width=1\textwidth,keepaspectratio]{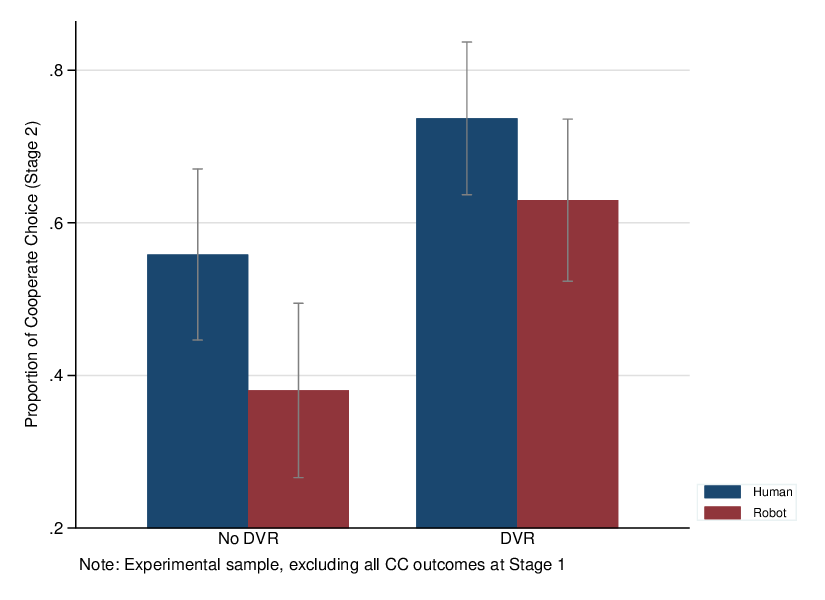}
    \caption{Choice of Cooperation at Stage 2 by experimental condition and partner type}
    \label{fig:barchoice}
\end{figure}

On average, all subjects eligible for treatment who proceeded to Stage 2 chose to Cooperate 58\% of the times, as shown in the top panel in Table \ref{tab:sumstat}. The breakdown by treatment and experimental condition is illustrated in Figure \ref{fig:barchoice} and summarized in Table \ref{tab:probCoop} in \ref{app:addtab}.

As Table \ref{tab:probCoop} shows, in the Control group, where no DVR is assigned to either the Human or Robot partner, the average cooperation rate is lower in the Robot condition.

On the other hand, in the Treatment group subjects facing either a Human or a Robotic partner who implement a DVR, display higher cooperation rates. Overall, this descriptive pattern suggests that in the absence of any verbal reaction, following the observed outcome of Stage 1, subjects are more likely to cooperate when facing a Human Partner: however, if a verbal reaction (DVR) is delivered by either a Robotic or Human partner, subjects are more likely to Cooperate than in the case of no-reaction. These patterns are visually summarized by Figure \ref{fig:barchoice}, in which the bars' height is clearly higher in the DVR group for both the Human and Robot conditions.

Our main hypotheses are tested through the estimation of Eq. \ref{eq:2}, whose outcome is shown in Table \ref{tab:interaction}. In this table there are three main coefficients of interest: Robot, DVR and DVR$\times$Robot.

\subsection{The effect of Partner type}
The estimated coefficient of Robot in Table \ref{tab:interaction} provides evidence that subjects display different behavior depending on the type of partner: being assigned to a Robot significantly decreases the cooperation rate, with an estimated probability between 16.5\% and 18\%, depending on specifications, as shown in the ``Margins'' columns. 
This result holds once control variables are included, as shown in Table \ref{tab:interactionCV} in \ref{app:addtab}, and the magnitude of the coefficient is substantially the same, implying a lower probability of cooperating, at between 15.6\% and 18.3\%. Overall, these results show that, \textit{ceteris paribus}, cooperation is strongly and significantly affected by partner type. This result is aligned with the evidence provided by a systematic review of experiments involving computer players performed by \citet{march2019behavioral} who finds that players' behavior differs across human vs.computer opponents; and that subjects generically behave more selfish and more rational when interacting with computers.

\begin{table}[htbp]\centering
\def\sym#1{\ifmmode^{#1}\else\(^{#1}\)\fi}
\caption{Main results: the effect of DVR and Robot, experimental sample}
\begin{adjustbox}{width=1\textwidth}
\centering
\begin{threeparttable}
\begin{tabular}{l*{6}{c}}
\toprule
                    &   \multicolumn{3}{c}{Benchmark}           &\multicolumn{3}{c}{Including Choice at Stage 1} \\ \cmidrule(lr){2-4} \cmidrule(lr){5-7}
Choice: ``Cooperate'' &   Full sample   &Excl. failed tests\tnote{$\dagger$}   &     Margins   &   Full sample   &Excl. failed tests\tnote{$\dagger$}   &     Margins   \\
\midrule
DVR$\times$Robot   &       0.117   &       0.297   &      0.028   &     -0.017   &       0.233   &    -0.004   \\
                    &     (0.503)   &     (0.545)   &     (0.122)   &     (0.525)   &     (0.569)   &     (0.127)   \\
                    &               &               &               &               &               &               \\
DVR            &       0.843   &       0.776   &       0.204   &       0.926   &       0.793   &       0.223   \\
                    &     (0.363)** &     (0.401)*  &    (0.088)** &     (0.378)** &     (0.419)*  &    (0.091)** \\
                    &               &               &               &               &               &               \\
Robot               &      -0.681   &      -0.874   &      -0.165   &      -0.762   &      -1.049   &      -0.184   \\
                    &     (0.355)*  &     (0.385)** &    (0.086)*  &     (0.370)** &     (0.406)***&    (0.089)** \\
                    &               &               &               &               &               &               \\
Cooperate (Stage 1) &               &               &               &       1.331   &       1.427   &       0.321   \\
                    &               &               &               &     (0.288)***&     (0.312)***&    (0.069)***\\
                    &               &               &               &               &               &               \\
Inst. order         &         Yes   &         Yes   &         Yes   &         Yes   &         Yes   &         Yes   \\
Wave       &         Yes   &         Yes   &         Yes   &         Yes   &         Yes   &         Yes   \\

Exp. day            &         Yes   &         Yes   &         Yes   &         Yes   &         Yes   &         Yes   \\
\midrule
Pseudo R-sq.        &        0.07   &        0.09   &               &        0.13   &        0.15   &               \\
Obs                 &         305   &         268   &         305   &         305   &         268   &         305   \\
LL                  &        -192   &        -167   &               &        -181   &        -156   &               \\
AIC                 &         416   &         366   &              &         395   &         346   &              \\
BIC                 &         476   &         424   &             &         459   &         407   &             \\
\bottomrule
\end{tabular}
\begin{tablenotes}
\item  \textit{Notes.} Logit model, dependent variable: choice of Cooperation at Stage 2, Prisoner's Dilemma. Subjects yielding the outcome ``CC'' in Stage 1 are excluded from this sample. Standard errors are in parentheses.
\item[$\dagger$] This sub-sample excludes all subjects who failed the test assessing their full comprehension of the instructions.
\item \sym{*} \(p<0.10\), \sym{**} \(p<0.05\), \sym{***} \(p<0.01\)
\end{tablenotes}
\end{threeparttable}
\end{adjustbox}
\label{tab:interaction}
\end{table}

\subsection{The effect of DVR (\textit{Dialogic Verbal Reaction})} 
The coefficient of DVR provides evidence to support the assertion that a DVR positively affects cooperation rates: in all models, subjects cooperate more when a DVR is introduced, with a probability of between 20.4\% and 22.3\%. In this case also, the results hold true once control variables are included, as shown in Table \ref{tab:interactionCV} in \ref{app:addtab}. All types of DVR exert a positive influence on the cooperation rate, as summarized by Figure \ref{fig:bar_DVR4}. However, the lack of statistical power, due to the limited sub-sample size, hinders a disaggregated analysis. In aggregate, this result provides new evidence of the positive effect of communication in fostering cooperation in strategic decision-making, strengthening and confirming previous similar findings \citep[such as][]{sally1995conversation, kollock1998social}. 

\subsection{Heterogeneity of DVR effect across partner types} 
\label{sub:H3}
In the previous section we showed that the effect of a DVR is strong and significant under both (Human and Robot) experimental conditions. In this section we test whether these effects are heterogeneous across partner types, i.e. whether being faced with a Human or Robot partner affects the way verbal interactions promote cooperative choices.
In Table \ref{tab:interaction}, the coefficient of the interaction variable $DVR\times Robot$ provides the outcome of this test.\footnote{\citet{ai2003interaction} show some concerns about the actual interpretation of interaction effects in nonlinear models, such as Logit. For this reason, we replicate all our main results through a Linear Probability Model, estimated through Ordinary Least Squares (OLS). These replications are shown in Appendix \ref{app:lpm}. All results are confirmed.} As the table shows, the coefficient is never significantly different from 0. 

Note that the coefficients of both $Robot$ and $DVR$ are significantly different from 0, as shown in the previous sections. Therefore, while the effect of the DVR is large and significant, this effect is not heterogeneous across partner types. In other words, once the respondent is randomly assigned to the DVR treatment group, he/she is more likely to cooperate with his/her partner, irrespective of whether the partner is a Human or a Robot.
The lack of heterogeneity provides a striking result: as long as the partner in the PD provides a DVR, subjects respond by increasing, on average, their probability of cooperating. Therefore, the ability to provide a verbal dialogic interaction - which indirectly evokes a commonly shared ethical norm in terms of the social desirability of cooperation - eliminates any previous differential cooperative attitudes with respect to human versus robots. These results are robust to the inclusion of the usual set of control variables, (as shown in Table \ref{tab:interactionCV}, in \ref{app:addtab}). 

The results shown in Table \ref{tab:interaction} may depend on the belief that a robot which is able to implement a sensible contextual dialogical interaction\footnote{Note that we can exclude that the effect is due to the mere fact of NAO being able to talk since it greets any subject it is interacting with, whether or not, in a later moment, it will perform a DVR.} is ``more human'', thus it elicits a behavior that a subject normally reserves for fellow human partners. Alternatively, the DVR may produce the effect by simply acting as a mere soft reminder of the social desirability of cooperation.\footnote{Similarly to the role played by the mentioning of the Ten Commandments as in \citep{mazar2008dishonesty}, or the honor code as in \citep{mccabe1993academic} in stimulating academic honesty.} 

Our experimental framework is unable to disentangle between these two competing and/or complementary explanations. However, it is worth noting that
NAO, the robot we used in the experiment, despite being able to display appropriate gaze and body gesture cues to increase its appearance of ``socialness'', is still far away to reproduce the pitch, accent and expressiveness of a natural human voice.\footnote{NAO communicated with the subject via a ``\textit{Wizard of Oz} system'' controlled by the laptop PC located in the control room. All moves and speech items were written and coded in the system. The robot followed a pre-programmed protocol where the experimenter did not need to speak or type anything during the interaction but only press a button to start the interaction as in \cite{laban2021tell}.} 

Further research is needed to disentangle these two possible explanation. One possible strategy would be to have the DVR provided by the Game Director (the pre-recorded voice over) commenting on the realization of socially sub-optimal results. In this case, the ``ethical reminder'' would be separated from the partner (and, in particular, from the robot). Nonetheless, our results seem consistent with some recent empirical findings support the importance of communication in HRI using the same social robots employed by our experiment (NAO). \citet{pelikan2016nao} find that subjects tend to use the same signals as in human–human interactions, such as adjusting word selection, turn length and prosody, thus adapting to the perceived limited capacity of the robot. \citet{tahir2018user} asses two modalities to deliver the feedback: audio only and audio combined with gestures and show that when audio and gesture are combined subjects better understand the feedback delivered by the robot. 

\section{Robustness checks}
\label{sec:robustness}

In this section we provide robustness checks for the main result shown in this paper: while being matched with a robot partner decreases, \textit{ceteris paribus}, cooperation rates and receiving a DVR from a partner (of any type) increases them, the effect of a DVR is not heterogeneous across partner types, i.e. it does not depend on the partner being a robot or a human.

We address potential sources of bias in our results that may be driven by the presence of Econ students (as defined below); the subject's psychological and behavioral traits; the choice made by the partner and observed by the subject in stage 1; the subject's perception of robots' behavior; and the occurrence of gender biases. We address each one of these concerns in separate subsections and finally also check for possible differential outcomes when using sample splits.\footnote{All our main results hold also implementing a difference-in-difference estimation strategy. Results are available upon request.}

\subsection{Excluding Econ students}
To assess the robustness of our results, we carry out our econometric analysis on different sub-samples. First of all, we analyze the potential bias induced by students studying for BAs in Economics, Management, Finance and Banking (henceforth: Econ students). Empirical evidence shows that students in this population are more self-interested than their non-Econ peers either because of a self-selection process \citep{frey2005selfish, carter1991economists} or because of an indoctrination effect \citep{frank1993does, haucap2014economists}.\footnote{A few studies \citep{yezer1996does, hu2003altruism} using different incentivized situations, find evidence that Econ students are more willing to cooperate.} Analysis of the causes of this differential behaviors lies beyond the scope of this project; nevertheless, we tested whether our results are driven by the inclusion of a substantial number of Econ students, who accounts for 36\% of the sample.\footnote{This percentage closely matches the proportion (37\%) of Econ Students on the total student population of the Milan Campus of the Catholic University.} 
Table \ref{tab:reactionNOecon} displays the results of the estimation of Eq. (2), as does Section \ref{sub:H3} where Econ students are excluded from the analysis. The main outcome of the analysis is confirmed, since the coefficient of the interacted variables is still not significant, and both the size and significance of the DVR and Robot coefficient is consistent with the results shown in Section \ref{sub:H3}. To complete our analysis, in Table \ref{tab:reactionNOeconCV} in \ref{app:addtab} we show results of the same models when control variables are included. Also in this case, the main outcome of our analysis still hold. 

\begin{table}[h]\centering
\def\sym#1{\ifmmode^{#1}\else\(^{#1}\)\fi}
\caption{Robustness check: excluding  Econ students}
\begin{adjustbox}{width=1\textwidth}
\centering
\begin{threeparttable}
\begin{tabular}{l*{6}{c}}
\toprule
               &   \multicolumn{3}{c}{Benchmark}           &\multicolumn{3}{c}{Including Choice at Stage 1} \\ \cmidrule(lr){2-4} \cmidrule(lr){5-7}
Choice: ``Cooperate'' &   Full sample   &Excl. failed tests\tnote{$\dagger$}   &     Margins   &   Full sample   &Excl. failed tests\tnote{$\dagger$}   &     Margins   \\
\midrule
DVR$\times$Robot     &      0.016   &       0.420   &     0.004   &     -0.084   &       0.453   &     -0.019   \\
                    &     (0.664)   &     (0.732)   &     (0.154)   &     (0.690)   &     (0.762)   &     (0.157)   \\
                    &               &               &               &               &               &               \\
DVR            &       1.134   &       0.975   &       0.262   &       1.163   &       0.900   &       0.265   \\
                    &     (0.472)** &     (0.528)*  &     (0.108)** &     (0.495)** &     (0.554)   &     (0.111)** \\
                    &               &               &               &               &               &               \\
Robot               &      -0.648   &      -0.956   &      -0.150   &      -0.825   &      -1.308   &      -0.188   \\
                    &     (0.475)   &     (0.524)*  &     (0.110)   &     (0.488)*  &     (0.555)** &     (0.111)*  \\
                    &               &               &               &               &               &               \\

Cooperate (Stage 1) &               &               &               &       1.380   &       1.573   &       0.315   \\
                    &               &               &               &     (0.409)***&     (0.466)***&    (0.093)***\\
                    &               &               &               &               &               &               \\
Inst. order         &         Yes   &         Yes   &         Yes   &         Yes   &         Yes   &         Yes   \\
Wave         &         Yes   &         Yes   &         Yes   &         Yes   &         Yes   &         Yes   \\
Exp. day         &         Yes   &         Yes   &         Yes   &         Yes   &         Yes   &         Yes   \\
\midrule
Pseudo R-sq.        &        0.12   &        0.14   &               &        0.17   &        0.20   &               \\
Obs                 &         192   &         166   &         192   &         192   &         166   &         192   \\
LL                  &        -112   &         -95   &               &        -106   &         -89   &               \\
AIC                 &         257   &         223   &             &         246   &         212   &              \\
BIC                 &         309   &         272   &             &         302   &         265   &             \\
\bottomrule
\end{tabular}
\begin{tablenotes}
\item  \textit{Notes.} Logit model, dependent variable: choice of Cooperation at Stage 2, Prisoner's Dilemma. Students studying for BAs in Economics, Management, Finance and Banking are excluded from the sample. Subjects yielding the outcome ``CC'' in Stage 1 are excluded from this sample. Standard errors are in parentheses.
\item[$\dagger$] This sub-sample excludes all subjects who failed the test assessing their full comprehension of the instructions.
\item \sym{*} \(p<0.10\), \sym{**} \(p<0.05\), \sym{***} \(p<0.01\)
\end{tablenotes}
\end{threeparttable}
\end{adjustbox}
\label{tab:reactionNOecon}
\end{table}

\subsection{Controlling for psychological and behavioral factors}
As a further robustness check we assess whether our results are affected by underlying unobserved psychological and behavioral traits of the subjects. The choice to cooperate entails the risk of being taken advantage of and may be driven by the level of general trust towards other people that is ``embedded'' in each individual subject. Furthermore, greater or lesser degree of impulsiveness may drive the cooperation/non cooperation choice. Finally, cooperation may be related to subjects' personal levels of pure altruism. 

While the empirical literature has not reached a consensus on the relation between risk aversion and cooperation \citep{dolbear1966risk,sabater2002accounting,burks2009cognitive,proto2019intelligence}, trust is generally associated with a positive influence on cooperation \citep{tedeschi1969trust,cook2003experimental,jung2012trust}, while IDR and impulsiveness have been shown to be negatively correlated to cooperation \citep{yi2005relationship,streich2007time,jones2009delay,locey2013social,malesza2020effects}

For this reason, Table \ref{tab:reactionPSYCHO} shows the outcome of the same models estimated in Section \ref{sub:H3} by controlling for a set of psychological scales based on subjects' answers in Phase 1 (before the lab experimental session took place). These scales include a Risk Scale, retrieved from the General Risk Scale developed by \citet{dohmen2011individual}, a Trust Scale, retrieved from the Trust in People Scale \citep{election1964}, a simple 12-items matching measure of Intertemporal Discount Rate (IDR) based on \cite{thaler1981some} and a ``raw'' measure of generosity (Donation) proxied by the decision to donate 1 euro of their final reward to a charity of their choice at the end of the experiment.

As Table \ref{tab:reactionPSYCHO} shows, only the coefficient of the Trust scale is weakly significant and consistent with the current empirical evidence. All the other psychological and behavioral controls are not statistically significant and the estimated coefficients relating to the main effects are unchanged with respect to the main model.

\begin{table}[h]\centering
\def\sym#1{\ifmmode^{#1}\else\(^{#1}\)\fi}
\caption{Robustness check: controlling for psychological and behavioral factors}
\begin{adjustbox}{width=1\textwidth}
\centering
\begin{threeparttable}
\begin{tabular}{l*{6}{c}}
\toprule
                    &   \multicolumn{3}{c}{Benchmark}           &\multicolumn{3}{c}{Including Choice at Stage 1} \\ \cmidrule(lr){2-4} \cmidrule(lr){5-7}
Choice: ``Cooperate'' &   Full sample   &Excl. failed tests\tnote{$\dagger$}   &     Margins   &   Full sample   &Excl. failed tests\tnote{$\dagger$}   &     Margins   \\
\midrule
DVR$\times$Robot         &       0.167   &       0.310   &      0.040   &      0.035   &       0.213   &     0.008   \\
                    &     (0.500)   &     (0.535)   &     (0.121)   &     (0.520)   &     (0.557)   &     (0.126)   \\
                    &               &               &               &               &               &               \\
DVR                 &       0.836   &       0.787   &       0.202   &       0.928   &       0.854   &       0.224   \\
                    &     (0.359)** &     (0.391)** &    (0.087)** &     (0.375)** &     (0.409)** &    (0.090)** \\
                    &               &               &               &               &               &               \\
Robot               &      -0.726   &      -0.834   &      -0.176   &      -0.798   &      -0.981   &      -0.193   \\
                    &     (0.347)** &     (0.375)** &    (0.084)** &     (0.359)** &     (0.392)** &    (0.087)** \\
                    &               &               &               &               &               &               \\
Cooperate (Stage 1) &               &               &               &       1.226   &       1.296   &       0.296   \\
                    &               &               &               &     (0.276)***&     (0.296)***&    (0.067)***\\
                    &               &               &               &               &               &               \\
\textit{Psychological and behavioral controls:} &&&&&& \\
Risk scale			&     0.009   &     0.010   &     0.002   &     -0.013   &    -0.007   &    -0.003   \\
                    &    (0.057)   &    (0.060)   &    (0.014)   &    (0.060)   &    (0.063)   &    (0.014)   \\
                    &               &               &               &               &               &               \\
Trust scale         &       0.235   &       0.214   &      0.057   &       0.216   &       0.212   &      0.0522   \\
                    &     (0.135)*  &     (0.144)   &    (0.033)*  &     (0.140)   &     (0.150)   &    (0.034)   \\
                    &               &               &               &               &               &               \\
IDR                 &    -0.004   &    -0.002   &   -0.001   &    -0.003   &    -0.001  &   -0.001   \\
                    &   (0.006)   &   (0.006)   &   (0.001)   &   (0.006)   &   (0.006)   &   (0.001)   \\
                    &               &               &               &               &               &               \\
Donation                &       0.137   &      0.016   &      0.033   &      0.096   &     -0.036  &      0.023   \\
                    &     (0.250)   &     (0.269)   &    (0.061)   &     (0.259)   &     (0.279)   &    (0.063)   \\
                    &               &               &               &               &               &               \\
Inst. order         &         Yes   &         Yes   &         Yes   &         Yes   &         Yes   &         Yes   \\
Wave                &         Yes   &         Yes   &         Yes   &         Yes   &         Yes   &         Yes   \\
Exp. day            &         Yes   &         Yes   &         Yes   &         Yes   &         Yes   &         Yes   \\
\midrule
Pseudo R-sq.        &        0.07   &        0.08   &               &        0.12   &        0.13   &               \\
Obs                 &         305   &         268   &         305   &         305   &         268   &         305   \\
LL                  &        -192   &        -169   &               &        -182   &        -159   &               \\
AIC                 &         408   &         362   &              &         390   &         344   &              \\
BIC                 &         453   &         405   &              &         438   &         390   &              \\
\bottomrule
\end{tabular}
\begin{tablenotes}
\item  \textit{Notes.} Logit model, dependent variable: choice of Cooperation at Stage 2, Prisoner's Dilemma. Subjects yielding the outcome ``CC'' in Stage 1 are excluded from this sample. Standard errors are in parentheses.
\item[$\dagger$] This sub-sample excludes all subjects who failed the test assessing their full comprehension of the instructions.
\item \sym{*} \(p<0.10\), \sym{**} \(p<0.05\), \sym{***} \(p<0.01\)
\end{tablenotes}
\end{threeparttable}
\end{adjustbox}
\label{tab:reactionPSYCHO}
\end{table}

\subsection{Controlling for the partner's choice}
\label{sub:choicepartner}
We also assess the robustness of our results to the inclusion of partner's choice\footnote{On average, partners exhibit an overall cooperation rate of about 50\% at stage 1 with no significant difference across partner's types (the Chi-sq. test yields a $Prob.= 0.302$ in the full sample; $Prob.= 0.669$ in the Experimental sample).} in the online phase of the experiment. Table \ref{tab:reactionPCHOICE} summarizes the outcome of the analysis, showing that partners' cooperative behavior significantly increases the subjects' probability of cooperation at stage 2 only when previous subjects' choices are not included in the model (see columns 1 to 3 in Table \ref{tab:reactionPCHOICE}). Therefore, the partner's behavior is not affecting \textit{per se} the subsequent decision to cooperate, while partner type and occurrence of DVR are.
These results are unchanged when control variables are included, as shown in Table \ref{tab:reactionPCHOICEcv}, reported in Appendix \ref{app:addtab}.

\begin{table}[h]\centering
\def\sym#1{\ifmmode^{#1}\else\(^{#1}\)\fi}
\caption{Robustness check: controlling for Partner's choice}
\begin{adjustbox}{width=1\textwidth}
\centering
\begin{threeparttable}
\begin{tabular}{l*{9}{c}}
\toprule
                    &   \multicolumn{3}{c}{Partner's Choice at Stage 1}      &\multicolumn{3}{c}{Subject's Choice at Stage 1}     &\multicolumn{3}{c}{Subject+Partner's Choice at Stage 1}  \\ \cmidrule(lr){2-4} \cmidrule(lr){5-7} \cmidrule(lr){8-10}
Choice: ``Cooperate'' &   Full sample   &Excl. failed tests\tnote{$\dagger$}   &     Margins   &   Full sample   &Excl. failed tests\tnote{$\dagger$}   &     Margins  &   Full sample   &Excl. failed tests\tnote{$\dagger$}   &     Margins \\
\midrule
DVR$\times$Robot           &       0.154   &       0.355   &      0.0373   &     -0.017  &       0.233   &    -0.004   &     0.007   &       0.246   &     0.002   \\
                    &     (0.514)   &     (0.557)   &     (0.125)   &     (0.525)   &     (0.569)   &     (0.127)   &     (0.526)   &     (0.570)   &     (0.127)   \\
                    &               &               &               &               &               &               &               &               &               \\
DVR                 &       0.841   &       0.736   &       0.204   &       0.926   &       0.793   &       0.223   &       0.916   &       0.786   &       0.221   \\
                    &     (0.371)** &     (0.411)*  &    (0.090)** &     (0.378)** &     (0.419)*  &    (0.091)** &     (0.379)** &     (0.419)*  &    (0.091)** \\
                    &               &               &               &               &               &               &               &               &               \\
Robot               &      -0.743   &      -0.980   &      -0.180   &      -0.762   &      -1.049   &      -0.184   &      -0.766   &      -1.052   &      -0.185   \\
                    &     (0.363)** &     (0.396)** &    (0.088)** &     (0.370)** &     (0.406)***&    (0.089)** &     (0.370)** &     (0.407)***&    (0.090)** \\
                    &               &               &               &               &               &               &               &               &               \\
Partner Cooperates (Stage 1)            &       1.199   &       1.172   &       0.290   &               &               &               &       0.291   &       0.172   &      0.070   \\
                &     (0.340)***&     (0.352)***&    (0.083)***&               &               &               &     (0.445)   &     (0.466)   &     (0.107)   \\
                    &               &               &               &               &               &               &               &               &               \\
Cooperate (Stage 1) &               &               &               &       1.331   &       1.427   &       0.321   &       1.175   &       1.330   &       0.284   \\
                    &               &               &               &     (0.288)***&     (0.312)***&    (0.069)***&     (0.372)***&     (0.407)***&    (0.090)***\\
                    &               &               &               &               &               &               &               &               &               \\
Inst. order         &         Yes   &         Yes   &         Yes   &         Yes   &         Yes   &         Yes   &         Yes   &         Yes   &         Yes   \\
Wave         &         Yes   &         Yes   &         Yes   &         Yes   &         Yes   &         Yes   &         Yes   &         Yes   &         Yes   \\
Exp. day            &         Yes   &         Yes   &         Yes   &         Yes   &         Yes   &         Yes   &         Yes   &         Yes   &         Yes   \\
\midrule
Pseudo R-sq.        &        0.11   &        0.12   &               &        0.13   &        0.15   &               &        0.13   &        0.15   &               \\
Obs                 &         305   &         268   &         305   &         305   &         268   &         305   &         305   &         268   &         305   \\
LL                  &        -186   &        -161   &               &        -181   &        -156   &               &        -180   &        -156   &               \\
AIC                 &         405   &         357   &           .   &         395   &         346   &           .   &         397   &         348   &           .   \\
BIC                 &         468   &         418   &           .   &         459   &         407   &           .   &         464   &         412   &           .   \\
\bottomrule
\bottomrule
\end{tabular}
\begin{tablenotes}
\item  \textit{Notes.} Logit model, dependent variable: choice of Cooperation at Stage 2, Prisoner's Dilemma. Subjects yielding the outcome ``CC'' in Stage 1 are excluded from this sample. Standard errors are in parentheses.
\item[$\dagger$] This sub-sample excludes all subjects who failed the test assessing their full comprehension of the instructions.
\item \sym{*} \(p<0.10\), \sym{**} \(p<0.05\), \sym{***} \(p<0.01\)
\end{tablenotes}
\end{threeparttable}
\end{adjustbox}
\label{tab:reactionPCHOICE}
\end{table}

\subsection{Controlling for perceptions about Robots }
\label{sub:beliefsrobot}
Facing a robotic partner in a PD is currently an uncommon situation for human subjects. Since we do not know \textit{a priori} how our subjects perceive a robot partner (especially in terms of preferences and/or beliefs), we decided to include a simple questionnaire during Phase 1 in which we asked subjects to identify their perception of robots. In this way we obtained information about whether subjects perceived a robot as an ``Adaptive'' device, or as a device simply controlled by a predetermined set of program lines.\footnote{The set of all possible answers is listed in Table \ref{tab:descvars}. Also note that the vast majority of subjects answered either ``Robots executes a fixed list of commands and operations'' or ``A robot adapts its behavior to  interaction with a human being'', thus supporting our binary coding.}

Table \ref{tab:reactionADAPTIVE} provides a test of the potential heterogeneity driven by the perception of robots as an adaptive tool. Interestingly, also in this case, the main outcome of our analysis is unchanged, with both the DVR and Robot variable remaining statistically significant, while all the interaction terms, capturing potential heterogeneity, are not significantly different from 0. 
As a further check, we inspect whether the effect of a DVR is still significant in the Robot subsample, as shown in Table \ref{tab:reaction_adaptiveROBOT}. The outcome shown in the Table strengthens our main result, confirming that the DVR is the main driver of cooperative behavior in our experiment.

\begin{table}[h]\centering
\def\sym#1{\ifmmode^{#1}\else\(^{#1}\)\fi}
\caption{Robustness check: controlling for heterogeneity driven by perceptions of robots' behavior}
\begin{adjustbox}{width=1\textwidth}
\centering
\begin{threeparttable}
\begin{tabular}{l*{6}{c}}
\toprule
                    &   \multicolumn{3}{c}{Benchmark}           &\multicolumn{3}{c}{Including Choice at Stage 1} \\ \cmidrule(lr){2-4} \cmidrule(lr){5-7}
Choice: ``Cooperate'' &   Full sample   &Excl. failed tests\tnote{$\dagger$}   &     Margins   &   Full sample   &Excl. failed tests\tnote{$\dagger$}   &     Margins   \\
\midrule
DVR$\times$Robot$\times$Adaptive     &      -0.847   &      -0.968   &      -0.205   &      -1.271   &      -1.260   &      -0.307   \\
                    &     (1.014)   &     (1.093)   &     (0.246)   &     (1.063)   &     (1.149)   &     (0.257)   \\
                    &               &               &               &               &               &               \\
DVR$\times$Robot    &       0.489   &       0.740   &       0.119   &       0.529   &       0.793   &       0.128   \\
                    &     (0.669)   &     (0.722)   &     (0.162)   &     (0.697)   &     (0.752)   &     (0.168)   \\
                    &               &               &               &               &               &               \\
DVR$\times$Adaptive          &      -0.104   &     -0.083   &     -0.0251   &       0.115   &     -0.011   &      0.028   \\
                    &     (0.727)   &     (0.799)   &     (0.176)   &     (0.762)   &     (0.843)   &     (0.184)   \\
                    &               &               &               &               &               &               \\
Adaptive$\times$Robot         &       0.954   &       1.021   &       0.231   &       0.947   &       0.844   &       0.228   \\
                    &     (0.711)   &     (0.785)   &     (0.172)   &     (0.737)   &     (0.823)   &     (0.178)   \\
                    &               &               &               &               &               &               \\
DVR                 &       0.885   &       0.805   &       0.215   &       0.889   &       0.808   &       0.215   \\
                    &     (0.476)*  &     (0.518)   &     (0.115)*  &     (0.493)*  &     (0.536)   &     (0.119)*  \\
                    &               &               &               &               &               &               \\
Robot               &      -1.080   &      -1.302   &      -0.262   &      -1.152   &      -1.406   &      -0.278   \\
                    &     (0.468)** &     (0.511)** &     (0.114)** &     (0.484)** &     (0.532)***&     (0.117)** \\
                    &               &               &               &               &               &               \\
Adaptive            &      -0.251   &      -0.314   &     -0.061   &      -0.267   &      -0.155   &     -0.065   \\
                    &     (0.488)   &     (0.556)   &     (0.118)   &     (0.509)   &     (0.591)   &     (0.123)   \\
                    &               &               &               &               &               &               \\
Cooperate (Stage 1) &               &               &               &       1.350   &       1.447   &       0.326   \\
                    &               &               &               &     (0.292)***&     (0.318)***&    (0.070)***\\
                    &               &               &               &               &               &               \\
Wave         &         Yes   &         Yes   &         Yes   &         Yes   &         Yes   &         Yes   \\
Inst. order         &         Yes   &         Yes   &         Yes   &         Yes   &         Yes   &         Yes   \\
Exp. day            &         Yes   &         Yes   &         Yes   &         Yes   &         Yes   &         Yes   \\
\midrule
Pseudo R-sq.        &        0.08   &        0.09   &               &        0.14   &        0.16   &               \\
Obs                 &         305   &         268   &         305   &         305   &         268   &         305   \\
LL                  &        -191   &        -166   &               &        -179   &        -155   &               \\
AIC                 &         421   &         371   &             &         400   &         351   &             \\
BIC                 &         496   &         443   &          &         479   &         426   &             \\
\bottomrule
\end{tabular}
\begin{tablenotes}
\item  \textit{Notes.} Logit model, dependent variable: choice of Cooperation at Stage 2, Prisoner's Dilemma. Subjects yielding the outcome ``CC'' in Stage 1 are excluded from this sample. Standard errors are in parentheses.
\item[$\dagger$] This sub-sample excludes all subjects who failed the test assessing their full comprehension of the instructions.
\item \sym{*} \(p<0.10\), \sym{**} \(p<0.05\), \sym{***} \(p<0.01\)
\end{tablenotes}
\end{threeparttable}
\end{adjustbox}
\label{tab:reactionADAPTIVE}
\end{table}

\begin{table}[h]\centering
\def\sym#1{\ifmmode^{#1}\else\(^{#1}\)\fi}
\caption{Robustness check: controlling for perceptions of robots' behavior, Robot sample}
\begin{adjustbox}{width=1\textwidth}
\centering
\begin{threeparttable}
\begin{tabular}{l*{6}{c}}
\toprule
                    &   \multicolumn{3}{c}{Benchmark}           &\multicolumn{3}{c}{Including Choice at Stage 1} \\ \cmidrule(lr){2-4} \cmidrule(lr){5-7}
Choice: ``Cooperate'' &   Full sample   &Excl. failed tests\tnote{$\dagger$}   &     Margins   &   Full sample   &Excl. failed tests\tnote{$\dagger$}   &     Margins   \\
\midrule
DVR$\times$Adaptive          &      -0.754   &      -0.918   &      -0.188   &      -1.033   &      -1.124   &      -0.258   \\
                    &     (0.745)   &     (0.790)   &     (0.186)   &     (0.793)   &     (0.840)   &     (0.198)   \\
                    &               &               &               &               &               &               \\
DVR                 &       1.407   &       1.568   &       0.352   &       1.554   &       1.682   &       0.388   \\
                    &     (0.501)***&     (0.539)***&     (0.125)***&     (0.541)***&     (0.581)***&     (0.135)***\\
                    &               &               &               &               &               &               \\
Adaptive            &       0.684   &       0.705   &       0.171   &       0.698   &       0.698   &       0.174   \\
                    &     (0.538)   &     (0.571)   &     (0.134)   &     (0.564)   &     (0.597)   &     (0.141)   \\
                    &               &               &               &               &               &               \\

Cooperate (Stage 1) &               &               &               &       1.610   &       1.543   &       0.402   \\
                    &               &               &               &     (0.463)***&     (0.486)***&     (0.116)***\\
                    &               &               &               &               &               &               \\
Wave        &         Yes   &         Yes   &         Yes   &         Yes   &         Yes   &         Yes   \\
Inst. order         &         Yes   &         Yes   &         Yes   &         Yes   &         Yes   &         Yes   \\
Exp. day            &         Yes   &         Yes   &         Yes   &         Yes   &         Yes   &         Yes   \\
\midrule
Pseudo R-sq.        &        0.15   &        0.16   &               &        0.21   &        0.22   &               \\
Obs                 &         152   &         140   &         152   &         152   &         140   &         152   \\
LL                  &         -90   &         -81   &               &         -83   &         -76   &               \\
AIC                 &         212   &         195   &             &         200   &         185   &              \\
BIC                 &         260   &         242   &          &         252   &         235   &             \\
\bottomrule
\end{tabular}
\begin{tablenotes}
\item  \textit{Notes.} Logit model, dependent variable: choice of Cooperation at Stage 2, Prisoner's Dilemma. Subjects yielding the outcome ``CC'' in Stage 1 are excluded from this sample. Standard errors are in parentheses.
\item[$\dagger$] This sub-sample excludes all subjects who failed the test assessing their full comprehension of the instructions.
\item \sym{*} \(p<0.10\), \sym{**} \(p<0.05\), \sym{***} \(p<0.01\)
\end{tablenotes}
\end{threeparttable}
\end{adjustbox}
\label{tab:reaction_adaptiveROBOT}
\end{table}

\subsection{Controlling for gender bias}
Since we enrolled both male and female PhD students for the role of the Human partner, a possible concern may arise relating to subjects' biased responses depending on the partner's gender.
For this reason, we perform a robustness check by including an interaction term between the partner's gender and the DVR, whose outcomes are presented in Table \ref{tab:interaction_partner}. Obviously, this analysis is only performed in the Human sub-sample. As the results in the Table show, the main outcome of our analysis is confirmed: while the DVR coefficient remains strongly significant, this is not the case for every interaction.\footnote{Table \ref{tab:OLSinteraction_partner} in Appendix \ref{app:lpm} reports similar results, with p-values for DVR only slightly above the 10\% threshold.}

\begin{table}[h]\centering
\def\sym#1{\ifmmode^{#1}\else\(^{#1}\)\fi}
\caption{Test for potential gender bias in PD interactions, experimental sample}
\begin{adjustbox}{width=1\textwidth}
\centering
\begin{threeparttable}
\begin{tabular}{l*{6}{c}}
\toprule
                    &   \multicolumn{3}{c}{Benchmark}           &\multicolumn{3}{c}{Including Choice at Stage 1} \\ \cmidrule(lr){2-4} \cmidrule(lr){5-7}
Choice: ``Cooperate'' &   Full sample   &Excl. failed tests\tnote{$\dagger$}   &     Margins   &   Full sample   &Excl. failed tests\tnote{$\dagger$}   &     Margins   \\
\midrule
DVR $\times$ Female $\times$ Female partner &      -1.933   &      -1.333   &               &      -0.316   &       0.507   &               \\
                    &     (2.062)   &     (2.124)   &               &     (2.217)   &     (2.327)   &               \\
                    &               &               &               &               &               &               \\
DVR $\times$ Female &      0.0578   &      -0.162   &               &      -1.179   &      -1.549   &               \\
                    &     (1.757)   &     (1.827)   &               &     (1.890)   &     (1.981)   &               \\
                    &               &               &               &               &               &               \\
DVR $\times$ Female partner &       1.289   &       0.745   &               &      0.0832   &      -0.558   &               \\
                    &     (1.823)   &     (1.868)   &               &     (1.939)   &     (2.008)   &               \\
                    &               &               &               &               &               &               \\
Female $\times$ Female partner &       2.627   &       2.091   &               &       1.889   &       1.125   &               \\
                    &     (1.494)*  &     (1.557)   &               &     (1.532)   &     (1.623)   &               \\
                    &               &               &               &               &               &               \\
DVR                 &       0.888   &       1.137   &       0.193   &       2.049   &       2.300   &       0.240   \\
                    &     (1.593)   &     (1.646)   &    (0.0860)** &     (1.703)   &     (1.762)   &    (0.0892)***\\
                    &               &               &               &               &               &               \\
Female              &      -1.548   &      -1.202   &      -0.107   &      -1.108   &      -0.578   &      -0.131   \\
                    &     (1.295)   &     (1.359)   &    (0.0951)   &     (1.312)   &     (1.382)   &    (0.0942)   \\
                    &               &               &               &               &               &               \\
Female partner      &      -2.128   &      -1.273   &     -0.0679   &      -1.607   &      -0.651   &     -0.0728   \\
                    &     (1.381)   &     (1.461)   &     (0.121)   &     (1.396)   &     (1.489)   &     (0.123)   \\
                    &               &               &               &               &               &               \\
Cooperate (Stage 1) &               &               &               &       1.700   &       1.895   &       0.366   \\
                    &               &               &               &     (0.465)***&     (0.512)***&    (0.0988)***\\
                    &               &               &               &               &               &               \\
Inst. order         &         Yes   &         Yes   &         Yes   &         Yes   &         Yes   &         Yes   \\
Wave                &         Yes   &         Yes   &         Yes   &         Yes   &         Yes   &         Yes   \\
Exp. day            &         Yes   &         Yes   &         Yes   &         Yes   &         Yes   &         Yes   \\
\midrule
Pseudo R-sq.        &        0.11   &        0.10   &               &        0.19   &        0.20   &               \\
Obs                 &         153   &         128   &         153   &         153   &         128   &         153   \\
LL                  &         -88   &         -75   &               &         -81   &         -67   &               \\
AIC                 &         216   &         189   &              &         204   &         175   &              \\
BIC                 &         277   &         246   &              &         267   &         235   &              \\
\bottomrule
\end{tabular}
\begin{tablenotes}
\item  \textit{Notes.} Logit model, dependent variable: choice of Cooperation at Stage 2, Prisoner's Dilemma. Subjects yielding the outcome ``CC'' in Stage 1 are excluded from this sample. Standard errors are in parentheses.
\item[$\dagger$] This sub-sample excludes all subjects who failed the test assessing their full comprehension of the instructions.
\item \sym{*} \(p<0.10\), \sym{**} \(p<0.05\), \sym{***} \(p<0.01\)
\end{tablenotes}
\end{threeparttable}
\end{adjustbox}
\label{tab:interaction_partner}
\end{table}

\subsection{Sample splits}
In this section we replicate a test of RQ1 and RQ2 with the use of sample splits: since our experimental design involves a 2$\times$2 treatment/condition matrix, we exploit sample splits to assess the robustness of our main results by testing RQ1 and RQ2 separately and independently.
Table \ref{tab:partnereffect} summarizes the results relating to the test of RQ1 (whether subjects behave differently depending on partner type) by treatment groups (i.e. in DVR and No DVR subsamples). The upper panel in Table \ref{tab:partnereffect} refers to the control group, where no DVR was delivered by the partner. In this case, being assigned to a Robot significantly decreases the cooperation rate, with an estimated probability of about 18\%, as shown in the ``Margins'' columns. A similar result can be found in the treated group, reported in the bottom panel, although in this case only when the choice made by the subject in Stage 1 is accounted for. In the Appendix, Table \ref{tab:partnereffectCV} shows that the result is robust to the inclusion of a set of control variables. Overall, the main results are confirmed: subjects tend to cooperate less when facing a robotic partner, especially when it does not perform a verbal reaction, thus suggesting that, when subjects face a robot (which does not verbally react to the game outcome), they are more likely to choose not to cooperate than if they were facing a fellow human being not delivering a DVR.

\begin{table}[htbp]\centering
\def\sym#1{\ifmmode^{#1}\else\(^{#1}\)\fi}
\caption{Effect of Partner type, by treatment condition, experimental sample}
\begin{adjustbox}{width=1\textwidth}
\centering
\begin{threeparttable}
\begin{tabular}{l*{6}{c}}
\toprule
\textbf{No DVR condition} &&&&&& \\  
                    &   \multicolumn{3}{c}{Benchmark}           &\multicolumn{3}{c}{Including Choice at Stage 1} \\ \cmidrule(lr){2-4} \cmidrule(lr){5-7}
Choice: ``Cooperate'' &   Full sample   &Excl. failed tests\tnote{$\dagger$}   &     Margins   &   Full sample   &Excl. failed tests\tnote{$\dagger$}   &     Margins   \\
\midrule
Robot               &      -0.710   &      -0.874   &      -0.177   &      -0.713   &      -0.931   &      -0.177   \\
                    &     (0.378)*  &     (0.408)** &    (0.0941)*  &     (0.387)*  &     (0.418)** &    (0.0963)*  \\
                    &               &               &               &               &               &               \\
Cooperate (Stage 1) &               &               &               &       1.131   &       1.012   &       0.281   \\
                    &               &               &               &     (0.429)***&     (0.453)** &     (0.106)***\\
                    &               &               &               &               &               &               \\
Inst. order         &         Yes   &         Yes   &         Yes   &         Yes   &         Yes   &         Yes   \\
Wave         &         Yes   &         Yes   &         Yes   &         Yes   &         Yes   &         Yes   \\
Exp. day            &         Yes   &         Yes   &         Yes   &         Yes   &         Yes   &         Yes   \\
\midrule
Pseudo R-sq.        &        0.08   &        0.10   &               &        0.12   &        0.13   &               \\
Obs                 &         148   &         127   &         148   &         148   &         127   &         148   \\
LL                  &         -94   &         -79   &               &         -90   &         -76   &               \\
AIC                 &         216   &         185   &          &         210   &         182   &             \\
BIC                 &         257   &         225   &          &         255   &         225   &            \\
\midrule
\textbf{DVR condition} &&&&&& \\
                    &   \multicolumn{3}{c}{Benchmark}           &\multicolumn{3}{c}{Including Choice at Stage 1} \\ \cmidrule(lr){2-4} \cmidrule(lr){5-7}
Choice: 'Cooperate' &   Full sample   &Excl. failed tests\tnote{$\dagger$}   &     Margins   &   Full sample   &Excl. failed tests\tnote{$\dagger$}   &     Margins   \\
\midrule
Robot               &      -0.504   &      -0.482   &      -0.106   &      -0.842   &      -0.892   &      -0.170   \\
                    &     (0.380)   &     (0.415)   &    (0.080)   &     (0.424)** &     (0.471)*  &    (0.085)** \\
                    &               &               &               &               &               &               \\
Cooperate (Stage 1) &               &               &               &       1.885   &       2.253   &       0.381   \\
                    &               &               &               &     (0.457)***&     (0.524)***&    (0.0907)***\\
                    &               &               &               &               &               &               \\
Inst. order         &         Yes   &         Yes   &         Yes   &         Yes   &         Yes   &         Yes   \\
Wave         &         Yes   &         Yes   &         Yes   &         Yes   &         Yes   &         Yes   \\
Exp. day            &         Yes   &         Yes   &         Yes   &         Yes   &         Yes   &         Yes   \\
\midrule
Pseudo R-sq.        &        0.07   &        0.09   &               &        0.17   &        0.22   &               \\
Obs                 &         157   &         141   &         157   &         157   &         141   &         157   \\
LL                  &         -91   &         -81   &               &         -82   &         -69   &               \\
AIC                 &         210   &         189   &              &         193   &         169   &            \\
BIC                 &         253   &         230   &              &         239   &         213   &              \\
\bottomrule
\end{tabular}
\begin{tablenotes}
\item  \textit{Notes.} Logit model, dependent variable: choice of Cooperation at Stage 2, Prisoner's Dilemma. Subjects yielding the outcome ``CC'' in Stage 1 are excluded from this sample. Standard errors are in parentheses.
\item[$\dagger$] This sub-sample excludes all subjects who failed the test assessing their full comprehension of the instructions.
\item \sym{*} \(p<0.10\), \sym{**} \(p<0.05\), \sym{***} \(p<0.01\)
\end{tablenotes}
\end{threeparttable}
\end{adjustbox}
\label{tab:partnereffect}
\end{table}

Table \ref{tab:dialog} provides a test of H2, directly addressing the effect of verbal reactions given by both human and robot partners. Also in this case, our test is performed through a sample split, in this case by partner types: by investigating the effect of the DVR within the same experimental condition (i.e. the partner type) we are able to investigate the pure effect of verbal interactions irrespective of partner type. 

\begin{table}[htbp]\centering
\def\sym#1{\ifmmode^{#1}\else\(^{#1}\)\fi}
\caption{Effect of Treatment, by partner type, experimental sample}
\begin{adjustbox}{width=1\textwidth}
\centering
\begin{threeparttable}
\begin{tabular}{l*{6}{c}}
\toprule
\textbf{Partner=Robot} &&&&&& \\  
                    &   \multicolumn{3}{c}{Benchmark}           &\multicolumn{3}{c}{Including Choice at Stage 1} \\ \cmidrule(lr){2-4} \cmidrule(lr){5-7}
Choice: ``Cooperate'' &   Full sample   &Excl. failed tests\tnote{$\dagger$}   &     Margins   &   Full sample   &Excl. failed tests\tnote{$\dagger$}   &     Margins   \\
\midrule
DVR            &       1.073   &       1.149   &       0.268   &       1.089   &       1.160   &       0.272   \\
                    &     (0.374)***&     (0.391)***&    (0.0933)***&     (0.394)***&     (0.411)***&    (0.0983)***\\
                    &               &               &               &               &               &               \\
Cooperate (Stage 1) &               &               &               &       1.570   &       1.512   &       0.392   \\
                    &               &               &               &     (0.456)***&     (0.481)***&     (0.114)***\\
                    &               &               &               &               &               &               \\
Inst. order         &         Yes   &         Yes   &         Yes   &         Yes   &         Yes   &         Yes   \\
Wave            &         Yes   &         Yes   &         Yes   &         Yes   &         Yes   &         Yes   \\
Exp. day            &         Yes   &         Yes   &         Yes   &         Yes   &         Yes   &         Yes   \\
\midrule
Pseudo R-sq.        &        0.14   &        0.15   &               &        0.20   &        0.21   &               \\
Obs                 &         152   &         140   &         152   &         152   &         140   &         152   \\
LL                  &         -91   &         -82   &               &         -84   &         -77   &               \\
AIC                 &         210   &         192   &              &         198   &         183   &              \\
BIC                 &         252   &         234   &              &         244   &         227   &             \\
\midrule
\textbf{Partner=Human} &&&&&& \\
                    &   \multicolumn{3}{c}{Benchmark}           &\multicolumn{3}{c}{Including Choice at Stage 1} \\ \cmidrule(lr){2-4} \cmidrule(lr){5-7}
Choice: 'Cooperate' &   Full sample   &Excl. failed tests\tnote{$\dagger$}   &     Margins   &   Full sample   &Excl. failed tests\tnote{$\dagger$}   &     Margins   \\
\midrule
DVR            &       0.937   &       0.977   &       0.209   &       1.131   &       1.130   &       0.246   \\
                    &     (0.379)** &     (0.425)** &    (0.0840)** &     (0.411)***&     (0.465)** &    (0.0881)***\\
                    &               &               &               &               &               &               \\
Cooperate (Stage 1) &               &               &               &       1.598   &       1.798   &       0.348   \\
                    &               &               &               &     (0.433)***&     (0.478)***&    (0.0930)***\\
                    &               &               &               &               &               &               \\
Inst. order         &         Yes   &         Yes   &         Yes   &         Yes   &         Yes   &         Yes   \\
Wave            &         Yes   &         Yes   &         Yes   &         Yes   &         Yes   &         Yes   \\
Exp. day            &         Yes   &         Yes   &         Yes   &         Yes   &         Yes   &         Yes   \\
\midrule
Pseudo R-sq.        &        0.08   &        0.07   &               &        0.15   &        0.17   &               \\
Obs                 &         153   &         128   &         153   &         153   &         128   &         153   \\
LL                  &         -92   &         -77   &               &         -84   &         -69   &               \\
AIC                 &         212   &         182   &          &         199   &         168   &             \\
BIC                 &         254   &         222   &             &         244   &         211   &             \\
\bottomrule
\end{tabular}
\begin{tablenotes}
\item  \textit{Notes.} Logit model, dependent variable: choice of Cooperation at Stage 2, Prisoner's Dilemma. Subjects yielding the outcome ``CC'' in Stage 1 are excluded from this sample. Standard errors are in parentheses.
\item[$\dagger$] This sub-sample excludes all subjects who failed the test assessing their full comprehension of the instructions.
\item \sym{*} \(p<0.10\), \sym{**} \(p<0.05\), \sym{***} \(p<0.01\)
\end{tablenotes}
\end{threeparttable}
\end{adjustbox}
\label{tab:dialog}
\end{table}

As the Table shows, the coefficient of the DVR is strongly positive and significant in both the Human and Robot experimental conditions. The upper panel in Table \ref{tab:dialog} shows that when the robot partner delivers a DVR, subjects are on average 27\% more likely to choose Cooperate in the PD than in the control group. A similar result can be found in the Human sub-sample, in which the probability of cooperation is between 21\% and 25\% higher when the human partner delivers a DVR than in the case he/she is not.\footnote{We also checked that this result is not affected by some sort of gender bias and found that no differential effect can be found depending on both partner's gender, subject's gender and the interaction between the two. A summary of this robustness check is reported in the Appendix, in Table \ref{tab:interaction_partner}, showing substantially no significant coefficients in the interaction terms.} 
It is worth noting that the effect is particularly strong when controlling for the outcome of Stage 1, i.e. the choice of respondents in the previous ``blind'' stage of the experiment. In other words, the result holds when taking account of the subject's cooperative attitude, which, as expected, is the most important predictor of his/her choices.

\clearpage

\section{Discussion and conclusions}
In a seminal paper, Farrel and Rabin state: ``People in reality do not seem to [...] question each other's statements as much, as game theory suggests they should'' \cite[p. 108]{farrell1996cheap}. Thus, people tend to respond even to cheap talks. Does what is true for Human-Human interactions, hold for Human-Robot interactions?

In this paper, we devised a randomized experiment in which human subjects are randomly matched to either a human or an anthropomorphic robot partner and asked to perform a repeated Prisoner’s Dilemma to investigate (i) whether subjects behave differently depending on the nature of their partner (human or robot); (ii) whether a Dialogic Verbal Reaction (DVR), which implicitly refers to cooperation as a socially desirable strategy, influenced the subject's subsequent choice; (iii) whether the effect caused by the DVR depended on the (human or robotic) nature of the partner.

Our results suggest that subjects tend to act more cooperatively with fellow human beings, rather than with robots; are influenced by a DVR towards a more cooperative strategy; and finally, that the effect of a DVR is strong enough to make a difference in behavior, based on the partner's type, insignificant. 

We are aware of the possible limitation of extending lab experiment results to the ``real world'' - as evidenced by \citet{rabin1993incorporating,levitt2007laboratory,levitt2008homo}; however we are convinced that our results may suggest interesting implications in a number of cases where robots are used to interact with - often fragile - human beings (Nursing Homes, Care Facilities, Hospitals, etc.).


The main result on the DVR may depend on the belief that a robot which is able to implement a dialogical interaction with the subject appears to be ``more human'', thus deserving a behavior similar to that reserved for fellow human partners.  
Further research is still needed to disentangle the effects of verbal interactions in HRI versus the reminder effect performed by the message they deliver.



\newpage
\bibliographystyle{chicago}
\bibliography{biblio_robot}


\clearpage
\appendix
\setcounter{section}{0}
\renewcommand{\thesection}{Appendix \Alph{section}}
\renewcommand{\thefigure}{\Alph{section}\arabic{figure}}
\setcounter{figure}{0}
\renewcommand{\thetable}{\Alph{section}\arabic{table}}
\setcounter{table}{0}

\clearpage

\section{Linear Probability Models (OLS)}
\label{app:lpm}
\setcounter{figure}{0}
\setcounter{table}{0}

This section presents a replication of the main results shown in Section \ref{sec:results} by adopting a Linear Probability Model, estimated through Ordinary Least Squares (OLS), in place of Logit models adopted in the main texts.

\begin{table}[htbp]\centering
\def\sym#1{\ifmmode^{#1}\else\(^{#1}\)\fi}
\caption{LPM: Effect of Partner type, by treatment condition, experimental sample}
\begin{adjustbox}{width=1\textwidth}
\centering
\begin{threeparttable}
\begin{tabular}{l*{4}{c}}
\toprule
\textbf{No DVR condition} &&&& \\
 &   \multicolumn{2}{c}{Benchmark}           &\multicolumn{2}{c}{Including Choice at Stage 1} \\ \cmidrule(lr){2-3} \cmidrule(lr){4-5}
Choice: 'Cooperate' &   Full sample   &Excl. failed tests\tnote{$\dagger$}   &   Full sample   &Excl. failed tests\tnote{$\dagger$}     \\
\midrule
Robot               &      -0.159   &      -0.196   &      -0.151   &      -0.200   \\
                    &    (0.091)*  &    (0.097)** &    (0.091)*  &    (0.095)** \\
                    &               &               &               &               \\
Cooperate (Stage 1) &               &               &       0.233   &       0.209   \\
                    &               &               &    (0.088)***&    (0.094)** \\
                    &               &               &               &               \\
Inst. order         &         Yes   &         Yes   &         Yes   &         Yes   \\
Wave                &         Yes   &         Yes   &         Yes   &         Yes   \\
Exp. day            &         Yes   &         Yes   &         Yes   &         Yes   \\
\midrule
Adj. R-sq.          &        0.02   &        0.03   &        0.06   &        0.06   \\
Obs                 &         148   &         127   &         148   &         127   \\
LL                  &         -99   &         -83   &         -95   &         -80   \\
AIC                 &         225   &         194   &         220   &         190   \\
BIC                 &         267   &         233   &         265   &         233   \\
\midrule
\textbf{DVR condition} &&&& \\
 &   \multicolumn{2}{c}{Benchmark}           &\multicolumn{2}{c}{Including Choice at Stage 1} \\ \cmidrule(lr){2-3} \cmidrule(lr){4-5}
Choice: 'Cooperate' &   Full sample   &Excl. failed tests\tnote{$\dagger$}   &   Full sample   &Excl. failed tests\tnote{$\dagger$}     \\
\midrule
Robot               &      -0.100   &     -0.098   &      -0.143   &      -0.151   \\
                    &    (0.080)    &    (0.085)    &    (0.078)*  &    (0.081)*  \\
                    &               &               &               &               \\
Cooperate (Stage 1) &               &               &       0.359   &       0.407   \\
                    &               &               &    (0.086)***&    (0.089)***\\
                    &               &               &               &               \\
Inst. order         &         Yes   &         Yes   &         Yes   &         Yes   \\
Wave                &         Yes   &         Yes   &         Yes   &         Yes   \\
Exp. day            &         Yes   &         Yes   &         Yes   &         Yes   \\
\midrule
Adj. R-sq.          &        0.01   &        0.03   &        0.12   &        0.17   \\
Obs                 &         157   &         141   &         157   &         141   \\
LL                  &         -96   &         -85   &         -85   &         -73   \\
AIC                 &         219   &         197   &         201   &         175   \\
BIC                 &         262   &         238   &         247   &         219   \\
\bottomrule
\end{tabular}
\begin{tablenotes}
\item  \textit{Notes.} LPM model (OLS), dependent variable: choice of Cooperation at Stage 2, Prisoner's Dilemma. Subjects yielding the outcome ``CC'' at Stage 1 are excluded from this sample. Robust standard errors are in parentheses.
\item[$\dagger$] This sub-sample excludes all subjects who failed the test to assessing their full comprehension of the instructions.
\item \sym{*} \(p<0.10\), \sym{**} \(p<0.05\), \sym{***} \(p<0.01\)
\end{tablenotes}
\end{threeparttable}
\end{adjustbox}
\label{tab:OLSpartnereffect}
\end{table}

\begin{table}[htbp]\centering
\def\sym#1{\ifmmode^{#1}\else\(^{#1}\)\fi}
\caption{LPM: Effect of Treatment, by partner type, experimental sample}
\begin{adjustbox}{width=1\textwidth}
\centering
\begin{threeparttable}
\begin{tabular}{l*{4}{c}}
\toprule
\textbf{Partner=Robot} &&&& \\  
 &   \multicolumn{2}{c}{Benchmark}           &\multicolumn{2}{c}{Including Choice at Stage 1} \\ \cmidrule(lr){2-3} \cmidrule(lr){4-5}
Choice: ``Cooperate'' &   Full sample   &Excl. failed tests\tnote{$\dagger$}     &   Full sample   &Excl. failed tests\tnote{$\dagger$}     \\
\midrule
DVR                 &       0.228   &       0.243   &       0.211   &       0.224   \\
                    &    (0.082)***&    (0.085)***&    (0.078)***&    (0.081)***\\
                    &               &               &               &               \\
Cooperate (Stage 1) &               &               &       0.308   &       0.286   \\
                    &               &               &    (0.087)***&    (0.091)***\\
                    &               &               &               &               \\
Inst. order         &         Yes   &         Yes   &         Yes   &         Yes   \\
Wave                &         Yes   &         Yes   &         Yes   &         Yes   \\   
Exp. day            &         Yes   &         Yes   &         Yes   &         Yes   \\
\midrule
Adj. R-sq.          &        0.10   &        0.11   &        0.17   &        0.17   \\
Obs                 &         152   &         140   &         152   &         140   \\
LL                  &         -96   &         -87   &         -89   &         -81   \\
AIC                 &         219   &         201   &         208   &         192   \\
BIC                 &         261   &         242   &         253   &         236   \\
\midrule
\textbf{Partner=Human} &&&& \\  
 &   \multicolumn{2}{c}{Benchmark}           &\multicolumn{2}{c}{Including Choice at Stage 1} \\ \cmidrule(lr){2-3} \cmidrule(lr){4-5}
Choice: ``Cooperate'' &   Full sample   &Excl. failed tests\tnote{$\dagger$}     &   Full sample   &Excl. failed tests\tnote{$\dagger$}     \\
\midrule
DVR                 &       0.200   &       0.206   &       0.207   &       0.205   \\
                    &    (0.080)** &    (0.089)** &    (0.079)***&    (0.088)** \\
                    &               &               &               &               \\
Cooperate (Stage 1) &               &               &       0.313   &       0.352   \\
                    &               &               &    (0.086)***&    (0.091)***\\
                    &               &               &               &               \\
Inst. order         &         Yes   &         Yes   &         Yes   &         Yes   \\
Wave                &         Yes   &         Yes   &         Yes   &         Yes   \\   
Exp. day            &         Yes   &         Yes   &         Yes   &         Yes   \\
\midrule
Adj. R-sq.          &        0.01   &       -0.01   &        0.10   &        0.10   \\
Obs                 &         153   &         128   &         153   &         128   \\
LL                  &         -97   &         -81   &         -89   &         -73   \\
AIC                 &         221   &         190   &         208   &         175   \\
BIC                 &         264   &         229   &         254   &         218   \\
\bottomrule
\end{tabular}
\begin{tablenotes}
\item  \textit{Notes.} LPM model (OLS), dependent variable: choice of Cooperation at Stage 2, Prisoner's Dilemma. Subjects yielding the outcome ``CC'' at Stage 1 are excluded from this sample. Robust standard errors are in parentheses.
\item[$\dagger$] This sub-sample excludes all subjects who failed the test to assessing their full comprehension of the instructions.
\item \sym{*} \(p<0.10\), \sym{**} \(p<0.05\), \sym{***} \(p<0.01\)
\end{tablenotes}
\end{threeparttable}
\end{adjustbox}
\label{tab:OLSdialog}
\end{table}

\begin{table}[htbp]\centering
\def\sym#1{\ifmmode^{#1}\else\(^{#1}\)\fi}
\caption{LPM: Heterogeneity of Treatment by partner type, experimental sample}
\begin{adjustbox}{width=1\textwidth}
\centering
\begin{threeparttable}
\begin{tabular}{l*{4}{c}}
\toprule
                    &   \multicolumn{2}{c}{Benchmark}           &\multicolumn{2}{c}{Including Choice at Stage 1} \\ \cmidrule(lr){2-3} \cmidrule(lr){4-5}
Choice: ``Cooperate'' &   Full sample   &Excl. failed tests\tnote{$\dagger$}    &   Full sample   &Excl. failed tests\tnote{$\dagger$}    \\
\midrule
DVR$\times$Robot          &      0.043   &      0.0843   &      0.016   &      0.066   \\
                    &     (0.114)   &     (0.123)   &     (0.111)   &     (0.118)   \\
                    &               &               &               &               \\
DVR                 &       0.184   &       0.166   &       0.184   &       0.156   \\
                    &    (0.081)** &    (0.089)*  &    (0.079)** &    (0.088)*  \\
                    &               &               &               &               \\
Robot               &      -0.162   &      -0.206   &      -0.167   &      -0.224   \\
                    &    (0.085)*  &    (0.091)** &    (0.084)** &    (0.090)** \\
                    &               &               &               &               \\
Cooperate (Stage 1) &               &               &       0.283   &       0.297   \\
                    &               &               &    (0.056)***&    (0.063)***\\
                    &               &               &               &               \\
Inst. order         &         Yes   &         Yes   &         Yes   &         Yes   \\
Wave                &         Yes   &         Yes   &         Yes   &         Yes   \\
Exp. day            &         Yes   &         Yes   &         Yes   &         Yes   \\
\midrule
Adj. R-sq.          &        0.05   &        0.06   &        0.12   &        0.13   \\
Obs                 &         305   &         268   &         305   &         268   \\
LL                  &        -202   &        -176   &        -190   &        -164   \\
AIC                 &         435   &         383   &         414   &         362   \\
BIC                 &         495   &         441   &         477   &         423   \\

\bottomrule
\end{tabular}
\begin{tablenotes}
\item  \textit{Notes.} LPM model (OLS), dependent variable: choice of Cooperation at Stage 2, Prisoner's Dilemma. Subjects yielding the outcome ``CC'' at Stage 1 are excluded from this sample. Robust standard errors are in parentheses.
\item[$\dagger$] This sub-sample excludes all subjects who failed the test to assessing their full comprehension of the instructions.
\item \sym{*} \(p<0.10\), \sym{**} \(p<0.05\), \sym{***} \(p<0.01\)
\end{tablenotes}
\end{threeparttable}
\end{adjustbox}
\label{tab:OLSinteraction}
\end{table}

\begin{table}[h]\centering
\def\sym#1{\ifmmode^{#1}\else\(^{#1}\)\fi}
\caption{LPM, Robustness check: excluding  Econ students}
\begin{adjustbox}{width=1\textwidth}
\centering
\begin{threeparttable}
\begin{tabular}{l*{4}{c}}
\toprule
                    &   \multicolumn{2}{c}{Benchmark}           &\multicolumn{2}{c}{Including Choice at Stage 1} \\ \cmidrule(lr){2-3} \cmidrule(lr){4-5}
Choice: ``Cooperate'' &   Full sample   &Excl. failed tests\tnote{$\dagger$}    &   Full sample   &Excl. failed tests\tnote{$\dagger$}    \\
\midrule
DVR$\times$Robot    &      0.034    &       0.124   &      0.028  &       0.138   \\
                    &     (0.145)   &     (0.156)   &     (0.140)   &     (0.152)   \\
                    &               &               &               &               \\
DVR                 &       0.217   &       0.179   &       0.199   &       0.143   \\
                    &    (0.094)**  &     (0.108)*  &    (0.095)** &     (0.109)   \\
                    &               &               &               &               \\
Robot               &      -0.147   &      -0.210   &      -0.173   &      -0.262   \\
                    &     (0.114)   &     (0.123)*  &     (0.114)   &     (0.123)** \\
                    &               &               &               &               \\
Cooperate (Stage 1) &               &               &       0.264   &       0.287   \\
                    &               &               &    (0.080)***&    (0.088)***\\
                    &               &               &               &               \\
Inst. order         &         Yes   &         Yes   &         Yes   &         Yes   \\
Wave                &         Yes   &         Yes   &         Yes   &         Yes   \\
Exp. day            &         Yes   &         Yes   &         Yes   &         Yes   \\
\midrule
Adj. R-sq.          &        0.08   &        0.09   &        0.13   &        0.15   \\
Obs                 &         192   &         166   &         192   &         166   \\
LL                  &        -118   &        -100   &        -112   &         -94   \\
AIC                 &         269   &         233   &         259   &         223   \\
BIC                 &         321   &         283   &         314   &         275   \\

\bottomrule
\end{tabular}
\begin{tablenotes}
\item  \textit{Notes.} LPM model (OLS), dependent variable: choice of Cooperation at Stage 2, Prisoner's Dilemma. Students studying for BAs in Economics, Management, Finance and Banking are excluded from the sample. Subjects yielding the outcome ``CC'' at Stage 1 are excluded from this sample. Robust standard errors are in parentheses.
\item[$\dagger$] This sub-sample excludes all subjects who failed the test to assessing their full comprehension of the instructions.
\item \sym{*} \(p<0.10\), \sym{**} \(p<0.05\), \sym{***} \(p<0.01\)
\end{tablenotes}
\end{threeparttable}
\end{adjustbox}
\label{tab:OLSreactionNOecon}
\end{table}

\begin{table}[h]\centering
\def\sym#1{\ifmmode^{#1}\else\(^{#1}\)\fi}
\caption{LPM, Robustness check: controlling for psychological and behavioral factors}
\begin{adjustbox}{width=1\textwidth}
\centering
\begin{threeparttable}
\begin{tabular}{l*{4}{c}}
\toprule
 &   \multicolumn{2}{c}{Benchmark}           &\multicolumn{2}{c}{Including Choice at Stage 1} \\ \cmidrule(lr){2-3} \cmidrule(lr){4-5}
Choice: ``Cooperate'' &   Full sample   &Excl. failed tests\tnote{$\dagger$}     &   Full sample   &Excl. failed tests\tnote{$\dagger$}     \\
\midrule
DVR$\times$Robot          &      0.053   &      0.095   &      0.023   &      0.072   \\
                    &     (0.115)   &     (0.123)   &     (0.112)   &     (0.119)   \\
                    &               &               &               &               \\
DVR                 &       0.176   &       0.157   &       0.179   &       0.152   \\
                    &    (0.081)** &    (0.089)*  &    (0.079)** &    (0.088)*  \\
                    &               &               &               &               \\
Robot               &      -0.170   &      -0.212   &      -0.172   &      -0.227   \\
                    &    (0.086)** &    (0.092)** &    (0.085)** &    (0.091)** \\
                    &               &               &               &               \\
\textit{Psychological and behavioral controls:} &&&& \\
Risk scale          &     0.001     &     0.002   &    -0.003   &    -0.002  \\
                    &    (0.013)    &    (0.014)   &    (0.012)   &    (0.013)   \\
                    &               &               &               &               \\
Trust scale         &      0.048   &      0.041   &      0.041   &      0.036   \\
                    &    (0.029)   &    (0.032)   &    (0.028)   &    (0.031)   \\
                    &               &               &               &               \\
IDR                 &   -0.001   &   -0.001   &   -0.001   &   -0.000   \\
                    &   (0.001)   &   (0.001)   &   (0.001)   &   (0.001)   \\
                    &               &               &               &               \\
Donation                &      0.037   &      0.015   &      0.027   &     0.008   \\
                    &    (0.058)   &    (0.063)   &    (0.056)   &    (0.060)   \\
                    &               &               &               &               \\
Cooperate (Stage 1) &               &               &       0.276   &       0.293   \\
                    &               &               &    (0.061)***&    (0.065)***\\
                    &               &               &               &               \\
Inst. order         &         Yes   &         Yes   &         Yes   &         Yes   \\
Wave                &         Yes   &         Yes   &         Yes   &         Yes   \\
Exp. day            &         Yes   &         Yes   &         Yes   &         Yes   \\
\midrule
Adj. R-sq.          &        0.05   &        0.05   &        0.11   &        0.13   \\
Obs                 &         305   &         268   &         305   &         268   \\
LL                  &        -200   &        -175   &        -189   &        -163   \\
AIC                 &         440   &         389   &         420   &         369   \\
BIC                 &         514   &         461   &         498   &         444   \\
\bottomrule
\end{tabular}
\begin{tablenotes}
\item  \textit{Notes.} LPM model (OLS), dependent variable: choice of Cooperation at Stage 2, Prisoner's Dilemma. Subjects yielding the outcome ``CC'' at Stage 1 are excluded from this sample. Robust standard errors are in parentheses.
\item[$\dagger$] This sub-sample excludes all subjects who failed the test to assessing their full comprehension of the instructions.
\item \sym{*} \(p<0.10\), \sym{**} \(p<0.05\), \sym{***} \(p<0.01\)
\end{tablenotes}
\end{threeparttable}
\end{adjustbox}
\label{tab:OLSreactionPSYCHO}
\end{table}

\begin{table}[h]\centering
\def\sym#1{\ifmmode^{#1}\else\(^{#1}\)\fi}
\caption{LPM, Robustness check: controlling for heterogeneity driven by perceptions of robots' behavior}
\begin{adjustbox}{width=1\textwidth}
\centering
\begin{threeparttable}
\begin{tabular}{l*{4}{c}}
\toprule
                    &   \multicolumn{2}{c}{Benchmark}           &\multicolumn{2}{c}{Including Choice at Stage 1} \\ \cmidrule(lr){2-3} \cmidrule(lr){4-5}
Choice: ``Cooperate'' &   Full sample   &Excl. failed tests\tnote{$\dagger$}    &   Full sample   &Excl. failed tests\tnote{$\dagger$}    \\
\midrule
DVR$\times$Adaptive &      -0.165   &      -0.189   &      -0.185   &      -0.197   \\
                    &     (0.167)   &     (0.176)   &     (0.159)   &     (0.168)   \\
                    &               &               &               &               \\
DVR                 &       0.298   &       0.325   &       0.290   &       0.309   \\
                    &     (0.105)***&     (0.109)***&     (0.101)***&     (0.105)***\\
                    &               &               &               &               \\
Adaptive            &       0.142   &       0.139   &       0.134   &       0.129   \\
                    &     (0.127)   &     (0.131)   &     (0.124)   &     (0.130)   \\
                    &               &               &               &               \\
Cooperate (Stage 1) &               &               &       0.309   &       0.286   \\
                    &               &               &    (0.087)***&    (0.090)***\\
                    &               &               &               &               \\
Inst. order         &         Yes   &         Yes   &         Yes   &         Yes   \\
Wave                &         Yes   &         Yes   &         Yes   &         Yes   \\
Exp. day            &         Yes   &         Yes   &         Yes   &         Yes   \\
\midrule
Adj. R-sq.          &        0.09   &        0.11   &        0.17   &        0.17   \\
Obs                 &         152   &         140   &         152   &         140   \\
LL                  &         -95   &         -86   &         -88   &         -80   \\
AIC                 &         221   &         203   &         210   &         194   \\
BIC                 &         270   &         251   &         261   &         244   \\

\bottomrule
\end{tabular}
\begin{tablenotes}
\item  \textit{Notes.} LPM model (OLS), dependent variable: choice of Cooperation at Stage 2, Prisoner's Dilemma. Subjects yielding the outcome ``CC'' at Stage 1 are excluded from this sample. Robust standard errors are in parentheses.
\item[$\dagger$] This sub-sample excludes all subjects who failed the test to assessing their full comprehension of the instructions.
\item \sym{*} \(p<0.10\), \sym{**} \(p<0.05\), \sym{***} \(p<0.01\)
\end{tablenotes}
\end{threeparttable}
\end{adjustbox}
\label{tab:OLSreactionADAPTIVE}
\end{table}

\begin{table}[h]\centering
\def\sym#1{\ifmmode^{#1}\else\(^{#1}\)\fi}
\caption{Test for potential gender bias in PD interactions, experimental sample}
\begin{adjustbox}{width=1\textwidth}
\centering
\begin{threeparttable}
\begin{tabular}{l*{4}{c}}
\toprule
 &   \multicolumn{2}{c}{Benchmark}           &\multicolumn{2}{c}{Including Choice at Stage 1} \\ \cmidrule(lr){2-3} \cmidrule(lr){4-5}
Choice: ``Cooperate'' &   Full sample   &Excl. failed tests\tnote{$\dagger$}     &   Full sample   &Excl. failed tests\tnote{$\dagger$}     \\
\midrule
DVR $\times$ Female $\times$ Female partner &     -0.450   &      -0.343   &      -0.233   &     -0.062   \\
                    &     (0.364)   &     (0.419)   &     (0.360)   &     (0.415)   \\
                    &               &               &               &               \\
DVR $\times$ Female &      0.066   &      0.025   &     -0.049   &      -0.152   \\
                    &     (0.292)   &     (0.353)   &     (0.284)   &     (0.345)   \\
                    &               &               &               &               \\
DVR $\times$ Female partner&       0.324   &       0.226   &       0.172   &      0.027   \\
                    &     (0.287)   &     (0.323)   &     (0.286)   &     (0.325)   \\
                    &               &               &               &               \\
Female $\times$ Female partner &      0.518   &       0.437   &       0.402   &       0.280   \\
                    &     (0.282)*  &     (0.325)   &     (0.283)   &     (0.335)   \\
                    &               &               &               &               \\
DVR                 &       0.132   &       0.178   &       0.218   &       0.301   \\
                    &     (0.226)   &     (0.268)   &     (0.222)   &     (0.269)   \\
                    &               &               &               &               \\
Female              &      -0.282   &      -0.237   &      -0.239   &      -0.152   \\
                    &     (0.230)   &     (0.270)   &     (0.227)   &     (0.276)   \\
                    &               &               &               &               \\
Female partner      &      -0.414   &      -0.272   &      -0.331   &      -0.178   \\
                    &     (0.245)*  &     (0.289)   &     (0.244)   &     (0.292)   \\
                    &               &               &               &               \\
Cooperate (Stage 1) &               &               &       0.309   &       0.353   \\
                    &               &               &    (0.086)***&    (0.092)***\\
                    &               &               &               &               \\
Inst. order         &         Yes   &         Yes   &         Yes   &         Yes   \\
Wave                &         Yes   &         Yes   &         Yes   &         Yes   \\
Exp. day            &         Yes   &         Yes   &         Yes   &         Yes   \\
\midrule
Adj. R-sq.          &        0.01   &       -0.03   &        0.09   &        0.09   \\
Obs                 &         153   &         128   &         153   &         128   \\
LL                  &         -93   &         -79   &         -86   &         -70   \\
AIC                 &         227   &         197   &         214   &         183   \\
BIC                 &         287   &         254   &         277   &         242   \\
\bottomrule
\end{tabular}
\begin{tablenotes}
\item  \textit{Notes.} LPM model (OLS), dependent variable: choice of Cooperation at Stage 2, Prisoner's Dilemma. Subjects yielding the outcome ``CC'' at Stage 1 are excluded from this sample. Robust standard errors are in parentheses.
\item[$\dagger$] This sub-sample excludes all subjects who failed the test to assessing their full comprehension of the instructions.
\item \sym{*} \(p<0.10\), \sym{**} \(p<0.05\), \sym{***} \(p<0.01\)
\end{tablenotes}
\end{threeparttable}
\end{adjustbox}
\label{tab:OLSinteraction_partner}
\end{table}

\clearpage

\section{Balance tests}
\label{app:balance}
\setcounter{figure}{0}
\setcounter{table}{0}

\begin{figure}[h]
    \centering
    \includegraphics[width=1\textwidth,keepaspectratio]{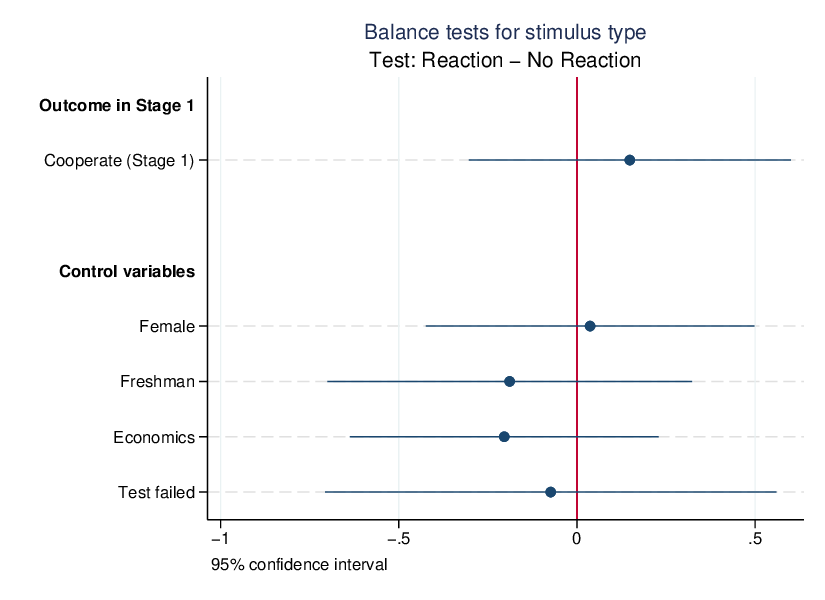}
    \caption{Summary of balance tests for assignment to treatment (DVR)}
    \label{fig:balanceTREAT}
\end{figure}

\begin{figure}
    \centering
    \includegraphics[width=1\textwidth,keepaspectratio]{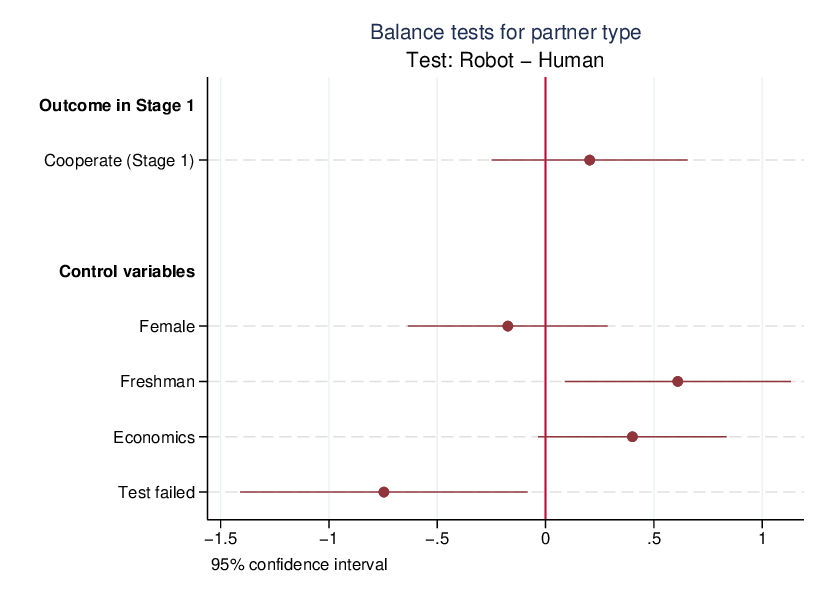}
    \caption{Summary of balance tests for assignment to Human/Robot condition}
    \label{fig:balancePARTNER}
\end{figure}

\begin{figure}
    \centering
    \includegraphics[width=1\textwidth,keepaspectratio]{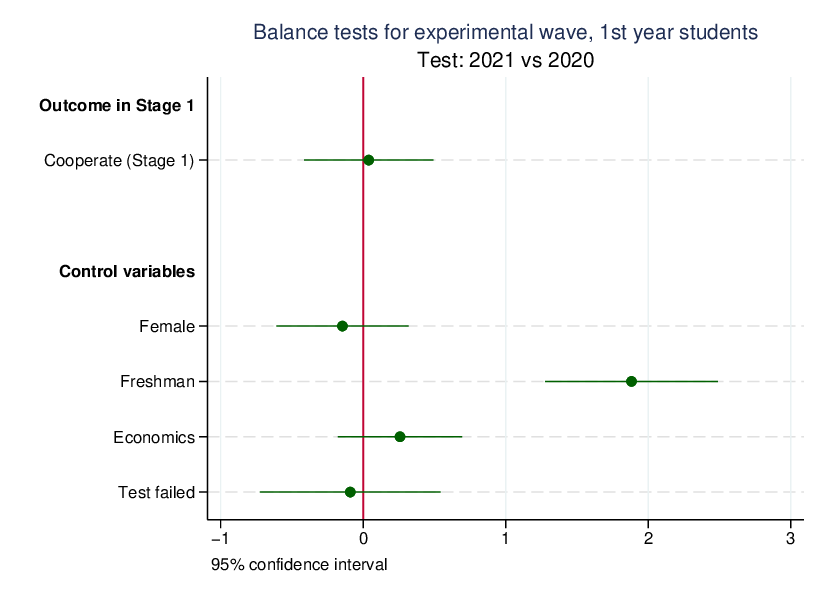}
    \caption{Summary of balance tests for experimental wave}
    \label{fig:balanceWAVE}
\end{figure}
\clearpage

\section{Additional tables}
\label{app:addtab}
\setcounter{figure}{0}
\setcounter{table}{0}

This section presents additional tables that are referenced in the main text but moved here for reasons of space.

\begin{table}[htbp]\centering
\def\sym#1{\ifmmode^{#1}\else\(^{#1}\)\fi}
\caption{Cooperation patterns at Stage 2, by Treatment group and Experimental condition}
\centering
\begin{threeparttable}
\begin{tabular}{l*{3}{r}}\toprule
 & Mean & St. Err. & Obs  \\
\midrule
\rule{0pt}{4ex}\textbf{No DVR group} &       &       &    \\
Partner=Robot & 0.380 & 0.058 & 71     \\
Partner=Human & 0.558 & 0.057 & 77   \\
\midrule
\rule{0pt}{4ex}\textbf{DVR group} &       &       &   \\
Partner=Robot & 0.630 & 0.054 & 81    \\
Partner=Human & 0.737 & 0.051 & 76     \\

\bottomrule
\end{tabular}
\end{threeparttable}
\label{tab:probCoop}
\end{table}

\begin{table}[]
\centering
\caption{Outline of \textit{Dialogic Verbal Reactions} performed by Partner}
\begin{threeparttable}
\begin{tabular}{clp{8cm}}
\toprule
\textbf{Outcome}    & \textbf{DVR type} & \textbf{Speech performed by partner} \\
(Subject, Partner)  & \\ \midrule
(C, N)                              & Apology & ``I realize I made a mistake in our online interaction. I meant to press C to cooperate. However I pressed N by mistake. I am really sorry! I will be more careful next time'' \\
&& \\
(N, C)                              & Reprimand  &  ``I am really upset. If you had chosen to cooperate we would have gained 6 euro each, a reasonable amount! On the contrary you exploited my goodwill and I got nothing'' \\ && \\
(N, N)                              & Disappointment & ``What a pity! If we had chosen to cooperate we would have gained 6 euros each. Why not cooperate in the next round?'' \\
&& \\
(C, C)                              & None     &     \\
\bottomrule
\end{tabular}
\label{tab:stimulus}
\end{threeparttable}
\end{table}

\begin{table}[htbp]\centering
\def\sym#1{\ifmmode^{#1}\else\(^{#1}\)\fi}
\caption{Variables description}
\begin{adjustbox}{height=0.48\textheight}
\centering
\begin{threeparttable}
\begin{tabular}{lp{12cm}}\toprule
                    &        Description  \\
\midrule
\textit{Outcome variables} & \\
Choice (Stage 2)    & Main outcome, binary variable equal to 1 if subject chooses ``Cooperate'' in PD's Stage 2, 0 otherwise \\
Choice (Stage 1)    & Binary variable equal to 1 if subject chooses ``Cooperate'' in PD's Stage 1, 0 otherwise \\
& \\
\textit{Treatment variables} & \\
DVR         &  Treatment variable, assuming value equal to 1 if partner delivers a \textit{Dialogic Verbal Reaction} after the outcome of Stage 1 \\
Robot       & Experimental condition, assuming value equal to 1 if subject is assigned to a Robot partner, 0 if he/she is assigned to a Human partner \\
& \\
\textit{Experiment-related controls} & \\
Instruction order   & Categorical variable recording the PD's outcome shown first to subject during the instruction session. Categories: ``CC'' (omitted reference category), ``CN'', ``NC'', ``NN'', where first letter refers to subjects' choice, second to partners'.  \\
Wave    & Binary variable, equal to 1 if experiment occurred in 2021, 0 in 2020 \\
Experimental day & Categorical variable recording the sequential number of each day of the experiment. Values from 1 (omitted reference category) for first day to 9 for ninth day \\
& \\
\textit{Individual characteristics} & \\
Female      &   Binary variable, equal to 1 if subject is Female, 0 otherwise \\
Freshman    &  Binary variable, equal to 1 if subject is in the first year of his/her BA, 0 if he/she is in second year \\
Econ        & Binary variable, equal to 1 if subject is enrolled in BAs in Economics, Management, Finance and Banking \\
Fail test       & Binary variable, equal to 1 if subject failed the pre-game test after the instruction session \\
&\\
\textit{Psychological scales and behaviors} & \\
Risk scale & General risk scale \citep{dohmen2011individual} \\
Trust scale & General trust scale \citep{election1964} \\
IDR & Inter-temporal discount rate, proxied by a question comparing a hypothetical immediate reward of €10 to a large delayed amount (larger values in IDR implies higher discount rates) \\
Donation    &  Binary variable, equal to 1 if the subject choose to donate 1 euro of the final reward to a charity of his/her choice, choosing between \textit{Med\'ecins Sans Fronti\`eres} or \textit{Greenpeace} \\
&\\
\textit{Other controls} & \\
Adaptive    &   Binary variable based on the answers given in the online questionnaire to the question: ``When interacting with a human being how does a robot behave?''. The variable is equal to 1 if the subject chose the following answer: ``A robot adapts its behavior to the interaction with the human being'; the variable is equal to 0 if the subject chose one of the following answers: ``A robot privileges its own interest over the interest of the human being'', ``A robot matches the interest of the human being to its own interest'', ``A robot exhibits a random behavior'', ``Robots execute a fixed list of commands and operations'' \\

\bottomrule
\end{tabular}
\end{threeparttable}
\end{adjustbox}
\label{tab:descvars}
\end{table}

\begin{table}[htbp]\centering
\def\sym#1{\ifmmode^{#1}\else\(^{#1}\)\fi}
\caption{Summary statistics, full sample}
\centering
\begin{threeparttable}
\begin{tabular}{l*{1}{ccc}}
\toprule
                    &        Mean&          St. Dev.&       Obs\\
\midrule
\textbf{Experimental sample}      &&& \\
\rule{0pt}{4ex}\textit{Outcome variables}&        &            &            \\
Cooperate (Stage 1) &       0.673&       0.470&         343\\
Cooperate (Stage 2) &       0.580&       0.494&         305\\
\rule{0pt}{4ex}\textit{Control variables}&        &            &            \\
Female              &       0.697&       0.460&         343\\
Freshman            &       0.778&       0.416&         343\\
Economics           &       0.397&       0.490&         343\\
Test failed         &       0.128&       0.335&         343\\\midrule

\textbf{Full sample}      &&& \\
\rule{0pt}{4ex}\textit{Outcome variables}&        &            &            \\
Cooperate (Stage 1) &       0.793&       0.406&         541\\
Cooperate (Stage 2) &       0.659&       0.474&         490\\
\rule{0pt}{4ex}\textit{Control variables}&        &            &            \\
Female              &       0.682&       0.466&         541\\
Freshman            &       0.786&       0.411&         541\\
Economics           &       0.381&       0.486&         541\\
Test failed         &       0.131&       0.338&         541\\

\bottomrule
\end{tabular}
\begin{tablenotes}
\footnotesize
\item \textit{Notes:} The Experimental sample includes only subjects whose outcomes at Stage 1 are relevant for Treatment/Control assignment, i.e. this sample excludes all subjects yielding ``CC'' as Stage 1's outcome; Full sample includes all subjects who participated in Phase 2 of the research. 
\end{tablenotes}
\end{threeparttable}
\label{tab:sumstat}
\end{table}

\begin{table}[htbp]\centering
\def\sym#1{\ifmmode^{#1}\else\(^{#1}\)\fi}
\caption{Effect of Partner type, by treatment condition, experimental sample, including control variables}
\begin{adjustbox}{width=1\textwidth}
\centering
\begin{threeparttable}
\begin{tabular}{l*{6}{c}}
\toprule

\bottomrule
\end{tabular}
\begin{tablenotes}
\item  \textit{Notes.} Logit model, dependent variable: choice of Cooperation at Stage 2, Prisoner's Dilemma. Subjects yielding the outcome ``CC'' at Stage 1 are excluded from this sample. Standard errors are in parentheses. Control variables include: Female, Freshman, Economics and Failed test.
\item[$\dagger$] This sub-sample excludes all subjects who failed the test assessing their full comprehension of the instructions.
\item \sym{*} \(p<0.10\), \sym{**} \(p<0.05\), \sym{***} \(p<0.01\)
\end{tablenotes}
\end{threeparttable}
\end{adjustbox}
\label{tab:partnereffectCV}
\end{table}

\begin{table}[h]\centering
\def\sym#1{\ifmmode^{#1}\else\(^{#1}\)\fi}
\caption{Heterogeneity of Treatment by partner type, experimental sample, including controls}
\begin{adjustbox}{width=1\textwidth}
\centering
\begin{threeparttable}
\begin{tabular}{l*{6}{c}}
\toprule
                    &   \multicolumn{3}{c}{Benchmark}           &\multicolumn{3}{c}{Including Choice at Stage 1} \\ \cmidrule(lr){2-4} \cmidrule(lr){5-7}
Choice: ``Cooperate'' &   Full sample   &Excl. failed tests\tnote{$\dagger$}   &     Margins   &   Full sample   &Excl. failed tests\tnote{$\dagger$}   &     Margins   \\
\midrule
DVR$\times$Robot &       0.134   &       0.357   &      0.032   &      0.0170   &       0.300   &     0.004   \\
            &     (0.520)   &     (0.563)   &     (0.126)   &     (0.540)   &     (0.586)   &     (0.130)   \\
            &               &               &               &               &               &               \\
DVR    &       0.832   &       0.737   &       0.202   &       0.908   &       0.759   &       0.219   \\
            &     (0.371)** &     (0.409)*  &    (0.087)** &     (0.385)** &     (0.426)*  &    (0.093)** \\
            &               &               &               &               &               &               \\
Robot       &      -0.643   &      -0.859   &      -0.156   &      -0.759   &      -1.049   &      -0.183   \\
            &     (0.364)*  &     (0.392)** &    (0.088)*  &     (0.377)** &     (0.412)** &    (0.091)** \\
Cooperate (Stage 1)&               &               &               &       1.300   &       1.402   &       0.314   \\
            &               &               &               &     (0.292)***&     (0.316)***&    (0.071)***\\
            &               &               &               &               &               &               \\
Controls    &         Yes   &         Yes   &         Yes   &         Yes   &         Yes   &         Yes   \\
Inst. order         &         Yes   &         Yes   &         Yes   &         Yes   &         Yes   &         Yes   \\
Wave       &         Yes   &         Yes   &         Yes   &         Yes   &         Yes   &         Yes   \\

Exp. day            &         Yes   &         Yes   &         Yes   &         Yes   &         Yes   &         Yes   \\
\midrule
Pseudo R-sq.&        0.08   &        0.09   &               &        0.13   &        0.15   &               \\
Obs         &         305   &         268   &         305   &         305   &         268   &         305   \\
LL          &        -191   &        -166   &               &        -180   &        -155   &               \\
AIC         &         422   &         370   &        &         403   &         351   &             \\
BIC         &         496   &         438   &          &         481   &         423   &            \\
\bottomrule
\end{tabular}
\begin{tablenotes}
\item  \textit{Notes.} Logit model, dependent variable: choice of Cooperation at Stage 2, Prisoner's Dilemma. Subjects yielding the outcome ``CC'' at Stage 1 are excluded from this sample. Standard errors are in parentheses. Control variables include: Female, Freshman, Economics and Failed test.
\item[$\dagger$] This sub-sample excludes all subjects who failed the test assessing their full comprehension of the instructions.
\item \sym{*} \(p<0.10\), \sym{**} \(p<0.05\), \sym{***} \(p<0.01\)
\end{tablenotes}
\end{threeparttable}
\end{adjustbox}
\label{tab:interactionCV}
\end{table}

\begin{table}[h]\centering
\def\sym#1{\ifmmode^{#1}\else\(^{#1}\)\fi}
\caption{Heterogeneity of Treatment by partner type, excluding Econ students, including control variables}
\begin{adjustbox}{width=1\textwidth}
\centering
\begin{threeparttable}
\begin{tabular}{l*{6}{c}}
\toprule
               &   \multicolumn{3}{c}{Benchmark}           &\multicolumn{3}{c}{Including Choice at Stage 1} \\ \cmidrule(lr){2-4} \cmidrule(lr){5-7}
Choice: ``Cooperate'' &   Full sample   &Excl. failed tests\tnote{$\dagger$}   &     Margins   &   Full sample   &Excl. failed tests\tnote{$\dagger$}   &     Margins   \\
\midrule
DVR$\times$Robot    &    -0.004   &       0.453   &    -0.001  &      -0.113   &       0.501   &     -0.026   \\
                    &     (0.691)   &     (0.756)   &     (0.159)   &     (0.717)   &     (0.792)   &     (0.163)   \\
                    &               &               &               &               &               &               \\
DVR            &       1.162   &       0.982   &       0.268   &       1.193   &       0.885   &       0.271   \\
                    &     (0.484)** &     (0.540)*  &     (0.111)** &     (0.507)** &     (0.569)   &     (0.113)** \\
                    &               &               &               &               &               &               \\
Robot               &      -0.650   &      -0.938   &      -0.150   &      -0.862   &      -1.302   &      -0.196   \\
                    &     (0.484)   &     (0.529)*  &     (0.112)   &     (0.497)*  &     (0.562)** &     (0.113)*  \\
                    &               &               &               &               &               &               \\
Cooperate (Stage 1) &               &               &               &       1.388   &       1.567   &       0.315   \\
                    &               &               &               &     (0.416)***&     (0.468)***&    (0.094)***\\
                    &               &               &               &               &               &               \\
Controls            &         Yes   &         Yes   &         Yes   &         Yes   &         Yes   &         Yes   \\
Inst. order         &         Yes   &         Yes   &         Yes   &         Yes   &         Yes   &         Yes   \\
Wave         &         Yes   &         Yes   &         Yes   &         Yes   &         Yes   &         Yes   \\
Exp. day         &         Yes   &         Yes   &         Yes   &         Yes   &         Yes   &         Yes   \\
\midrule
Pseudo R-sq.        &        0.13   &        0.14   &               &        0.18   &        0.20   &               \\
Obs                 &         192   &         166   &         192   &         192   &         166   &         192   \\
LL                  &        -111   &         -95   &               &        -105   &         -89   &               \\
AIC                 &         261   &         226   &            &         251   &         216   &             \\
BIC                 &         323   &         282   &           &         316   &         275   &             \\
\bottomrule
\end{tabular}
\begin{tablenotes}
\item  \textit{Notes.} Logit model, dependent variable: choice of Cooperation at Stage 2, Prisoner's Dilemma. Students studying for BAs in Economics, Management, Finance and Banking are excluded from the sample. Subjects yielding the outcome ``CC'' at Stage 1 are excluded from this sample. Standard errors are in parentheses. Control variables include: Female, Freshman, Economics and Failed test.
\item[$\dagger$] This sub-sample excludes all subjects who failed the test assessing their full comprehension of the instructions.
\item \sym{*} \(p<0.10\), \sym{**} \(p<0.05\), \sym{***} \(p<0.01\)
\end{tablenotes}
\end{threeparttable}
\end{adjustbox}
\label{tab:reactionNOeconCV}
\end{table}

\begin{table}[h]\centering
\def\sym#1{\ifmmode^{#1}\else\(^{#1}\)\fi}
\caption{Robustness check: controlling for Partner's choice}
\begin{adjustbox}{width=1\textwidth}
\centering
\begin{threeparttable}
\begin{tabular}{l*{9}{c}}
\toprule

\bottomrule
\end{tabular}
\begin{tablenotes}
\item  \textit{Notes.} Logit model, dependent variable: choice of Cooperation at Stage 2, Prisoner's Dilemma. Subjects yielding the outcome ``CC'' in Stage 1 are excluded from this sample. Standard errors are in parentheses. Control variables include: Female, Freshman, Economics and Failed test.
\item[$\dagger$] This sub-sample excludes all subjects who failed the test assessing their full comprehension of the instructions.
\item \sym{*} \(p<0.10\), \sym{**} \(p<0.05\), \sym{***} \(p<0.01\)
\end{tablenotes}
\end{threeparttable}
\end{adjustbox}
\label{tab:reactionPCHOICEcv}
\end{table}

\clearpage

\section{Additional figures}
\label{app:addfig}
\setcounter{figure}{0}
\setcounter{table}{0}

\begin{figure}[h]
    \centering
    \includegraphics[width=1\textwidth,keepaspectratio]{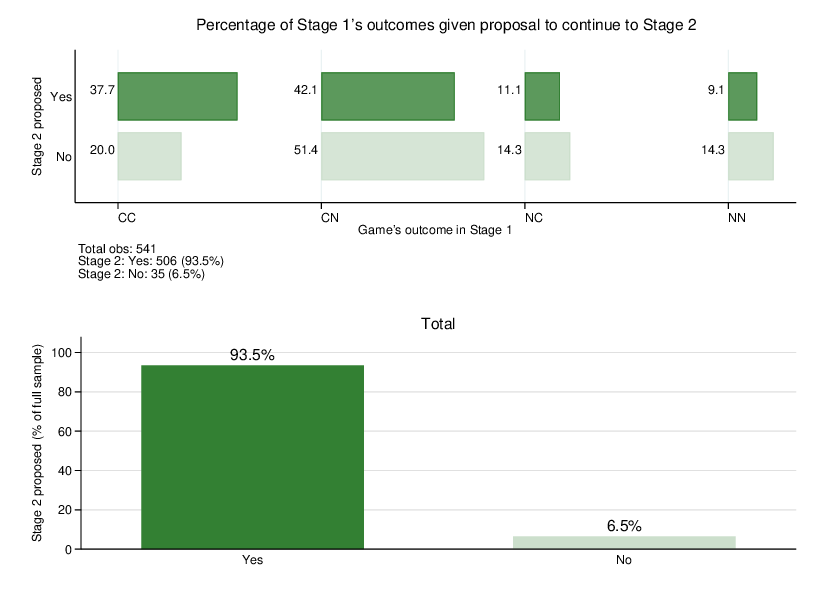}
    \caption{Percentage of respondents being proposed to move to Stage 2 by observed outcome in Stage 1.}
    \label{fig:bar_gostage2}
\end{figure}

\begin{figure}[h]
    \centering
    \includegraphics[width=1\textwidth,keepaspectratio]{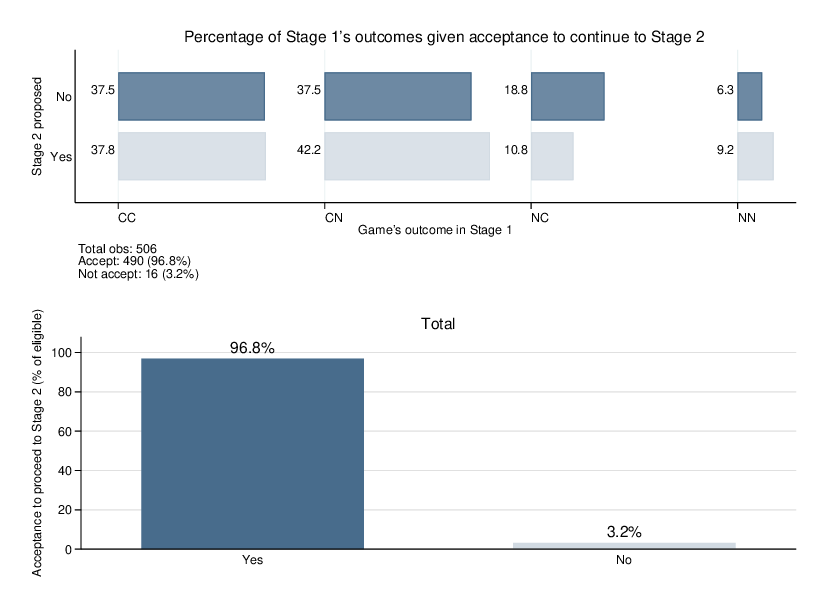}
    \caption{Percentage of respondents accepting to proceed to Stage 2 by observed outcome in Stage 1.}
    \label{fig:bar_acceptstage2}
\end{figure}

\begin{figure}[h]
    \centering
    \includegraphics[width=1\textwidth,keepaspectratio]{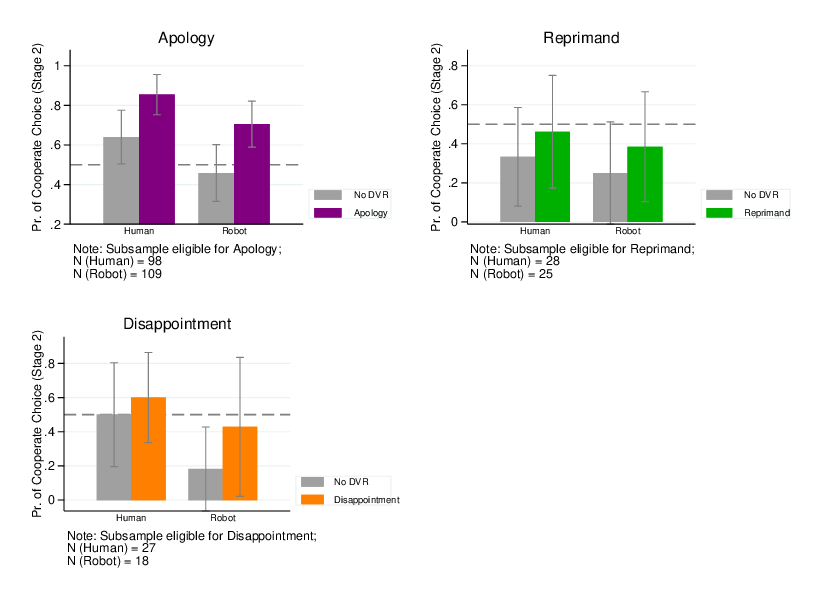}
    \caption{Choice of Cooperation in Stage 2 by experimental condition and type of DVR}
    \label{fig:bar_DVR4}
\end{figure}

\begin{figure}[h]
    \centering
    \includegraphics[width=0.5\textwidth,keepaspectratio]{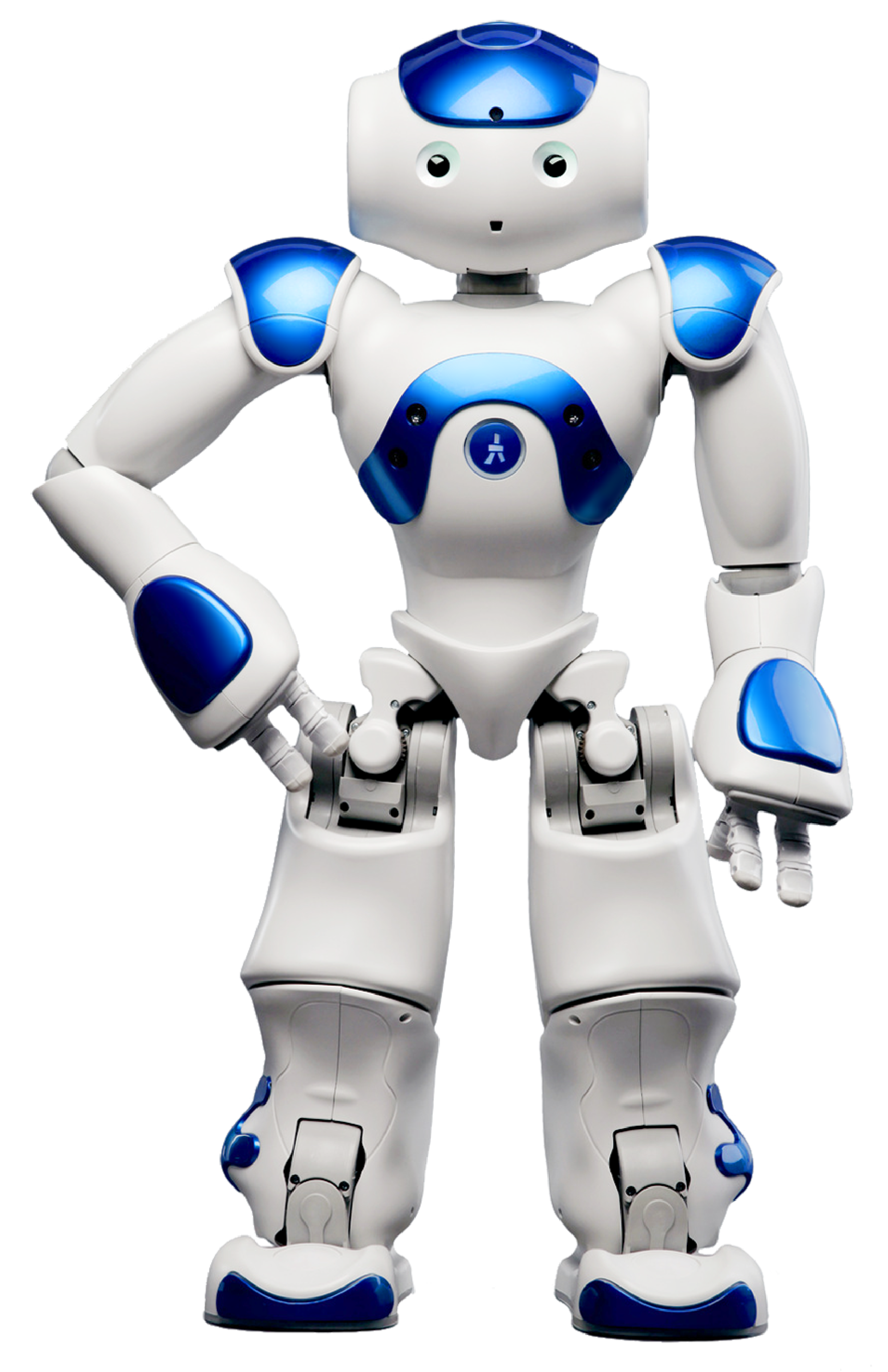}
    \caption{NAO robot, produced by Softbank Robotics}
    \label{fig:nao}
\end{figure}

\begin{figure}[h]
    \centering
    \includegraphics[width=1\textwidth,keepaspectratio]{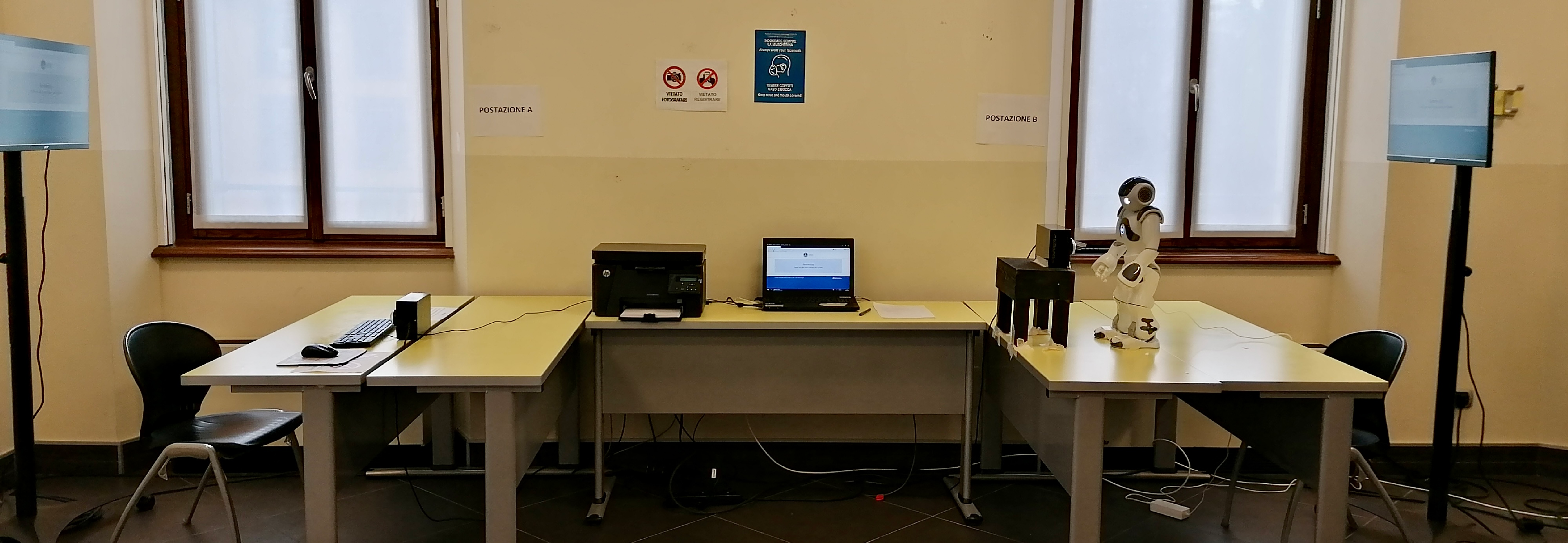}
    \caption{Experiment room}
    \label{fig:exproom}
\end{figure}

\end{document}